\documentstyle[12pt,amssym]{article}
\font\twelveof=msbm10 at 12pt
\def\R{\mbox{\twelveof R}}
\def\C{\mbox{\twelveof C}}
\def\Z{\mbox{\twelveof Z}}
\def\N{\mbox{\twelveof N}}

\begin{document}
\baselineskip=20pt plus 1pt minus1pt
\hfill ULB/229/CQ/99/1
\vspace{1.5cm}

\begin{center}

{\Large\bf\boldmath $C_{\lambda}$-extended oscillator algebras and some of their
deformations and applications to quantum mechanics}

\bigskip

{\large\bf C.\ Quesne\footnote{\rm Physique Nucl\'eaire Th\'eorique et Physique
Math\'ematique, Universit\'e Libre de Bruxelles, B-1050 Brussels, Belgium} and
N.\ Vansteenkiste\footnotemark[1]}

\end{center}

\vspace{10cm}
\noindent
Running head: $C_{\lambda}$-extended oscillator algebras

\noindent
Mailing address: C. Quesne, Physique Nucl\'eaire Th\'eorique et Physique
Math\'ematique, Universit\'e Libre de Bruxelles, Campus de la Plaine CP229,
Boulevard du Triomphe, B-1050 Brussels, Belgium

\noindent
Tel.: +32-2-6505559; fax: +32-2-6505045

\noindent
E-mail address: cquesne@ulb.ac.be
\newpage
%
%
\begin{center}
{\bf Abstract}
\end{center}

\noindent
$C_{\lambda}$-extended oscillator algebras generalizing the Calogero-Vasiliev
algebra, where $C_{\lambda}$ is the cyclic group of order $\lambda$, are studied
both from mathematical and applied viewpoints. Casimir operators of the algebras
are obtained, and used to provide a complete classification of their unitary
irreducible representations under the assumption that the number operator
spectrum is nondegenerate. Deformed algebras admitting Casimir operators
analogous to those of their undeformed counterparts are looked for,
yielding three
new algebraic structures. One of them includes the Brzezi\'nski {\em et al.}\
deformation of the Calogero-Vasiliev algebra as a special case. In its bosonic
Fock-space representation, the realization of $C_{\lambda}$-extended oscillator
algebras as generalized deformed oscillator ones is shown to provide a
bosonization of several variants of supersymmetric quantum mechanics:
parasupersymmetric quantum mechanics of order $p = \lambda-1$ for any
$\lambda$, as well as pseudosupersymmetric and orthosupersymmetric quantum
mechanics of order two for $\lambda=3$.
\newpage
%
%
\section{INTRODUCTION}
\label{sec:intro}
The oscillator algebra of creation, annihilation, and number operators plays a
central role in the investigation of many physical systems, and provides a
useful
tool in the theory of Lie algebra representations. Similarly, some of its
deformations (or extensions) have found applications to various physical
problems,
such as the description of systems with non-standard
statistics (Greenberg, 1990, 1991; Fivel, 1990; Meljanac {\em et al.}, 1994;
Meljanac and Milekovi\'c, 1996; Quesne, 1994a), the construction of integrable
lattice
models (Bogoliubov {\em et al.}, 1994), the investigation of nonlinearities in
quantum
optics (McDermott and Solomon, 1994; Solomon, 1998; Man'ko {\em et al.}, 1997),
the bosonization of supersymmetric quantum
mechanics (SSQM) (Bonatsos and Daskaloyannis, 1993a; Brzezi\'nski {\em et al.},
1993; Plyushchay, 1996a, b; Beckers {\em et al.}, 1997), as well as the
algebraic
treatment of quantum exactly solvable models (Daskaloyannis, 1992; Bonatsos and
Daskaloyannis, 1993b; Bonatsos {\em et al.}, 1993, 1994; Quesne, 1994b),
$n$-particle
integrable systems (Vasiliev, 1991; Polychron\-akos, 1992; Brink {\em et al.},
1992; Brink and Vasiliev, 1993; Quesne, 1995), pairing correlations in
nuclei (Bonatsos, 1992; Bonatsos and Daskaloyannis, 1992a), and vibrational
spectra of molecules (Chang {\em et al.}, 1991; Chang and Yan, 1991a, b, c;
Bonatsos and Daskaloyannis, 1992b, 1993c). In
addition, they have been used to construct representations of quantum universal
enveloping algebras of Lie algebras, also referred to as quantum
algebras (Biedenharn, 1989; Macfarlane, 1989; Sun and Fu, 1989; Hayashi, 1990;
Fairlie and Zachos, 1991; Fairlie and Nuyts, 1994).\par
%
%
Deformations of the oscillator algebra arose from successive
generalizations of the
Arik-Coon (Arik and Coon, 1976; Kuryshkin, 1980), and Biedenharn-Macfarlane
(Biedenharn, 1989; Macfarlane, 1989; Sun and Fu, 1989)
$q$-oscillators.
Various attempts have been made to introduce some order in the rich and varied
choice of deformed commutation relations by defining `generalized deformed
oscillator algebras' (GDOAs). Among them, one may quote the treatments due to
Jannussis {\em et al.} (1991), Jannussis (1993), Daskaloyannis (1991), Bonatsos
and Daskaloyannis (1993a), Irac-Astaud and Rideau (1992, 1993, 1994), McDermott
and Solomon (1994), Meljanac {\em et al.} (1994), Meljanac and Milekovi\'c
(1996),
Katriel and Quesne (1996), Quesne and
Vansteenkiste (1995, 1996, 1997). In the remainder of the present paper, we
shall refer to GDOAs as defined in the last references.\par
%
%
$G$-extended oscillator (or alternatively Heisenberg$^1$) algebras,
where $G$ is some finite group, appeared in connection with $n$-particle
integrable
models. It was shown (Vasiliev, 1991; Polychronakos, 1992; Brink {\em et al.},
1992; Brink and Vasiliev, 1993; Quesne, 1995) that they provide an algebraic
formulation of the Calogero model (Calogero,
1969a, b, 1971), or
some generalization thereof (Wolfes, 1974; Calogero and Marchioro, 1974). In the
former case, $G$ is the
symmetric
group $S_n$ (Polychronakos, 1992; Brink {\em et al.},
1992; Brink and Vasiliev, 1993). For two particles, the abelian group $S_2$ can
be
realized
in terms of Klein operator $K = (-1)^N$, where $N$ denotes the number
operator. The
$S_2$-extended oscillator algebra then becomes a GDOA, also known as the
Calogero-Vasiliev (Vasiliev, 1991), or modified (Brzezi\'nski {\em et al.},
1993) oscillator
algebra. Some
deformations of the latter have been extensively
studied (Brzezi\'nski {\em et al.},
1993; Macfarlane, 1994; Kosi\'nski {\em et al.}, 1997; Tsohantjis {\em et al.},
1997; Paolucci and Tsohantjis, 1997).\par
%
%
The purpose of the present paper is to study a new class of $G$-extended
oscillator
algebras (Quesne and Vansteenkiste, 1998), generalizing the one describing the
two-particle Calogero
model. Here $G$ is the cyclic group of order $\lambda$, $C_{\lambda} = \{I,
T, T^2,
\ldots, T^{\lambda-1}\}$, which for $\lambda=2$ is isomorphic to $S_2$. Such
$C_{\lambda}$-extended oscillator algebras
${\cal A}^{(\lambda)}_{\alpha_0 \alpha_1 \ldots \alpha_{\lambda-2}}$ have a rich
structure, since they depend upon $\lambda-1$ independent real parameters
$\alpha_0$, $\alpha_1$, $\ldots$,~$\alpha_{\lambda-2}$ (reducing to a single one
in the $\lambda=2$ case, corresponding to the $S_2$-extended oscillator
algebra).
Realizing $T$ in terms of the number operator $N$ converts ${\cal
A}^{(\lambda)}_{\alpha_0 \alpha_1 \ldots \alpha_{\lambda-2}}$ into a GDOA ${\cal
A}^{(\lambda)}(G(N))$.\par
%
%
The bosonic oscillator Hamiltonian $H_0$, associated with ${\cal
A}^{(\lambda)}(G(N))$, is equivalent to the two-particle Calogero
Hamiltonian for
$\lambda=2$, but exhibits entirely new features for $\lambda\ge3$ (Quesne and
Vansteenkiste, 1998). In
such a case, all the levels corresponding to a number of quanta equal to $\mu\,
\mbox{\rm mod}\,\lambda$ are equally spaced, but the ordering and spacing of
levels associated with different $\mu$ values depend on the algebra parameters
$\alpha_0$, $\alpha_1$, $\ldots$,~$\alpha_{\lambda-2}$. By appropriately
choosing
the latter, one may therefore obtain nondegenerate spectra, as well as spectra
exhibiting some $(\nu+1)$-fold degeneracies, where $\nu$ may take any value
in the
set $\{1, 2, \ldots, \lambda-1\}$.\par
%
%
The rich variety of spectra that may be obtained with $H_0$, as well as the
connection with the Calogero model for $\lambda=2$, makes it most likely that
some interesting applications will arise in one or another context. To help
towards
finding them, the construction of realizations of the ${\cal
A}^{(\lambda)}(G(N))$
generators in terms of differential operators is under current investigation and
will be reported elsewhere.\par
%
%
We may however already note that spectra that are a strictly equidistant
continuation of a triplet of `ground' states (which can be obtained here for
$\lambda=3$) arose in two studies of a class of potentials (with applications in
string theory) using either an advanced factorization method (Veselov and
Shabat, 1993), or a
nonlinear generalization of the Fock method (Eleonsky {\em et al.}, 1994, 1995;
Eleonsky and Korolev, 1995). Such spectra
can also
be obtained in SSQM by using cyclic shape invariant potentials of period
three (Sukhatme {\em et al.}, 1997). In this context, we recently showed that
three
appropriately
chosen ${\cal A}^{(3)}(G(N))$ algebras provide a matrix realization of
SSQM (Quesne and Vansteenkiste, 1999).\par
%
%
Another field wherein $C_{\lambda}$-extended oscillator algebras and their
deformations may be of interest is the study of coherent (or squeezed) states in
nonlinear quantum optics, wherein nonlinear oscillators are known to play an
important role (McDermott and Solomon, 1994; Solomon, 1998; Man'ko {\em et al.},
1997).\par
%
%
In the present paper, apart from studying some mathematical properties of
$C_{\lambda}$-extended oscillator algebras, we deal with some important
conceptual applications of these algebras. We indeed plan to show that they
provide
a bosonization (i.e., a realization in terms of only boson-like operators
without
fermion-like ones) of several variants of SSQM, namely parasupersymmetric
quantum mechanics (PSSQM) of  arbitrary order $p$ (Rubakov and Spiridonov, 1988;
Khare, 1992, 1993),
pseudoSSQM (Beckers {\em et al.}, 1995a, b; Beckers and Debergh, 1995a, b), and
orthosupersymmetric quantum mechanics (OSSQM)
of order two (Khare {\em et al.}, 1993a). These results generalize that
previously obtained for standard SSQM in terms of the Calogero-Vasiliev algebra
(Brzezi\'nski {\em et al.}, 1993; Plyushchay, 1996a, b).\par
%
%
In Section~\ref{sec:definition}, we review the definition of
$C_{\lambda}$-extended oscillator algebras, give their Casimir operators, and
present some of their realizations. In Section~\ref{sec:irreps}, we
classify their
unitary irreducible representations (unirreps). In Sections~\ref{sec:para},
\ref{sec:pseudo}, and
\ref{sec:ortho}, we consider their applications to PSSQM of arbitrary order
$p$, pseudoSSQM, and OSSQM of order two, respectively. In
Section~\ref{sec:deformations}, we construct some of their deformations.
Finally,
Section~\ref{sec:conclusion} contains the conclusion.\par
%
%
\section{\boldmath $C_{\lambda}$-EXTENDED OSCILLATOR ALGEBRAS}
\label{sec:definition}
A $C_{\lambda}$-extended oscillator algebra ${\cal A}^{(\lambda)}$, where
$\lambda$ may take any value in the set $\{2, 3, 4, \ldots\}$, is
defined (Quesne and Vansteenkiste, 1998) as an algebra generated by the
operators $I$,
$a^{\dagger}$, $a
= \left(a^{\dagger}\right)^{\dagger}$, $N = N^{\dagger}$, and
$T = \left(T^{\dagger}\right)^{-1}$, satisfying the relations
\begin{equation}
  \left[N, a^{\dagger}\right] = a^{\dagger}, \qquad [N, T] = 0, \qquad
T^{\lambda} = I, \label{eq:alg-def1-1}
\end{equation}
\begin{equation}
  \left[a, a^{\dagger}\right] = I + \sum_{\mu=1}^{\lambda-1} \kappa_{\mu}
T^{\mu},
  \qquad a^{\dagger} T = e^{- {\rm i}2\pi/\lambda} T a^{\dagger},
  \label{eq:alg-def1-2}
\end{equation}
together with their Hermitian conjugates. Here $\kappa_{\mu}$, $\mu = 1$,
2, $\ldots$,~$\lambda-1$, are some complex parameters restricted by the
conditions $\kappa_{\mu}^* = \kappa_{\lambda - \mu}$ (so that there remain
altogether $\lambda-1$ independent real parameters), and $T$ is the generator of
the cyclic group of order~$\lambda$, $C_{\lambda} = \{I, T, T^2, \ldots,
T^{\lambda-1}\}$ (or, more precisely, the generator of a unitary representation
thereof).\par
%
%
It is well known (Cornwell, 1984) that $C_{\lambda}$ has $\lambda$ inequivalent,
one-dimensional matrix unirreps $\Gamma^{\mu}$, $\mu = 0$, 1,
$\ldots$,~$\lambda-1$, which are such that
$\Gamma^{\mu}\left(T^{\nu}\right) = \exp({\rm i}2\pi \mu \nu/\lambda)$ for any
$\nu = 0$, 1, $\ldots$,~$\lambda-1$. The projection operator on the carrier
space
of~$\Gamma^{\mu}$ may be written as
\begin{equation}
  P_{\mu} = \frac{1}{\lambda} \sum_{\nu=0}^{\lambda-1} \Bigl(\Gamma^{\mu} \left(
  T^{\nu}\right) \Bigr)^* T^{\nu} = \frac{1}{\lambda} \sum_{\nu=0}^{\lambda-1}
  e^{-{\rm i}2\pi \mu\nu/\lambda}\, T^{\nu},   \label{eq:projector}
\end{equation}
and conversely $T^{\nu}$, $\nu=0$, 1, $\ldots$,~$\lambda-1$, may be expressed in
terms of the $P_{\mu}$'s as
\begin{equation}
  T^{\nu} = \sum_{\mu=0}^{\lambda-1}  e^{{\rm i}2\pi \mu\nu/\lambda} P_{\mu}.
  \label{eq:T}
\end{equation}
\par
%
%
The algebra defining relations~(\ref{eq:alg-def1-1}) and~(\ref{eq:alg-def1-2})
may therefore be rewritten in
terms of $I$, $a^{\dagger}$, $a$, $N$, and~$P_{\mu}^{\vphantom{\dagger}} =
P_{\mu}^{\dagger}$, $\mu=0$, 1, $\ldots$,~$\lambda-1$, as
\begin{equation}
  \left[N, a^{\dagger}\right] = a^{\dagger}, \qquad \left[N, P_{\mu}\right]
= 0, \qquad
  a^{\dagger} P_{\mu} = P_{\mu+1}\, a^{\dagger},  \label{eq:alg-def2-1}
\end{equation}
\begin{equation}
  \sum_{\mu=0}^{\lambda-1} P_{\mu} = I, \qquad P_{\mu} P_{\nu} =
\delta_{\mu,\nu}
  P_{\mu},  \label{eq:alg-def2-2}
\end{equation}
\begin{equation}
  \left[a, a^{\dagger}\right] = I + \sum_{\mu=0}^{\lambda-1} \alpha_{\mu}
P_{\mu},
  \label{eq:alg-def2-3}
\end{equation}
where we use the convention $P_{\mu'} = P_{\mu}$ if $\mu' - \mu = 0\, {\rm
mod}\,
\lambda$ (and similarly for other operators or parameters labelled by $\mu$,
$\mu'$).
Equations~(\ref{eq:alg-def2-1})--(\ref{eq:alg-def2-3}) depend
upon $\lambda$ real parameters
$\alpha_{\mu}$, $\mu = 0$, 1, $\ldots$,~$\lambda-1$, defined in terms of the
$\kappa_{\mu}$'s by
\begin{equation}
  \alpha_{\mu} = \sum_{\nu=1}^{\lambda-1} \exp({\rm i}2\pi \mu\nu/\lambda)
  \kappa_{\nu}, \qquad \mu=0, 1, \ldots, \lambda-1,
\end{equation}
and restricted by the condition
\begin{equation}
  \sum_{\mu=0}^{\lambda-1} \alpha_{\mu} = 0.  \label{eq:alpha-cond}
\end{equation}
Hence, we may eliminate one of them, e.g., $\alpha_{\lambda-1}$, and denote the
algebra by ${\cal A}^{(\lambda)}_{\alpha_0 \alpha_1 \ldots \alpha_{\lambda-2}}$.
It will, however, often prove convenient to work instead with the $\lambda$
dependent parameters $\alpha_0$, $\alpha_1$, $\ldots$,~$\alpha_{\lambda-1}$.\par
%
%
{}From Eqs.~(\ref{eq:alg-def1-1}) and~(\ref{eq:alg-def1-2}),
or~(\ref{eq:alg-def2-1})--(\ref{eq:alg-def2-3}), it is
easy to check
that ${\cal
A}^{(\lambda)}_{\alpha_0 \alpha_1 \ldots \alpha_{\lambda-2}}$ admits the
following Casimir operators:
\begin{equation}
  {\cal C}_1 = e^{{\rm i} 2\pi N},  \label{eq:casimir1}
\end{equation}
\begin{equation}
  {\cal C}_2 = e^{-{\rm i} 2\pi N/\lambda}\, T = \sum_{\mu=0}^{\lambda-1}
e^{-{\rm
i}
  2\pi (N-\mu)/\lambda} P_{\mu},  \label{eq:casimir2}
\end{equation}
\begin{equation}
  {\cal C}_3 = N + \sum_{\mu=0}^{\lambda-1} \beta_{\mu} P_{\mu} -
a^{\dagger} a,
  \label{eq:casimir3}
\end{equation}
where
\begin{equation}
  \beta_{\mu} = \sum_{\nu=0}^{\mu-1} \alpha_{\nu}, \qquad \mu = 1, 2, \ldots,
  \lambda-1,  \label{eq:beta}
\end{equation}
and $\beta_0 = \beta_{\lambda} = 0$. The first two operators are not
functionally
independent since
\begin{equation}
  {\cal C}_1 {\cal C}_2^{\lambda} = I.  \label{eq:casimir-rel}
\end{equation}
{}From Eq.~(\ref{eq:casimir2}), it follows that the cyclic group generator
$T$ can be
rewritten in terms of $N$ and ${\cal C}_2$ as
\begin{equation}
  T = e^{{\rm i} 2\pi N/\lambda} {\cal C}_2.  \label{eq:Tbis}
\end{equation}
\par
%
%
The simplest realization of the cyclic group $C_{\lambda}$ uses functions
of~$N$.
By taking ${\cal C}_2 = I$ in Eq.~(\ref{eq:Tbis}), and using
Eq.~(\ref{eq:projector}),
we obtain
\begin{equation}
  T = e^{{\rm i}2\pi N/\lambda}, \qquad P_{\mu} = \frac{1}{\lambda}
  \sum_{\nu=0}^{\lambda-1} e^{{\rm i}2\pi \nu (N-\mu)/\lambda}, \qquad \mu = 0,
 1, \ldots, \lambda-1.    \label{eq:T-GDOA}
\end{equation}
With such a choice, ${\cal A}^{(\lambda)}_{\alpha_0 \alpha_1 \ldots
\alpha_{\lambda-2}}$ becomes a GDOA ${\cal A}^{(\lambda)}(G(N))$, i.e., an
algebra
generated by $I$, $a^{\dagger}$ , $a = \left(a^{\dagger}\right)^{\dagger}$,
and $N =
N^{\dagger}$, subject to the relations
\begin{equation}
  \left[N, a^{\dagger}\right] = a^{\dagger}, \qquad  \left[a,
a^{\dagger}\right] =
  G(N),  \label{eq:GDOA-def}
\end{equation}
where $G(N)$ is some Hermitian, analytic function of $N$ (Quesne and
Vansteenkiste, 1995). In the
present case,
\begin{equation}
  G(N) = I + \sum_{\mu=0}^{\lambda-1} \alpha_{\mu} P_{\mu},  \label{eq:G}
\end{equation}
where $P_{\mu}$ is given by Eq.~(\ref{eq:T-GDOA}).\par
%
%
According to the GDOA general theory (see Quesne and Vansteenkiste (1995, 1996,
1997) and
references quoted therein), one may define a structure function $F(N)$,
which is the
solution of the difference equation $F(N+1) - F(N) = G(N)$ such that $F(0)
= 0$. For
$G(N)$ given in Eq.~(\ref{eq:G}), one finds
\begin{equation}
  F(N) = N + \sum_{\mu=0}^{\lambda-1} \beta_{\mu} P_{\mu},  \label{eq:F}
\end{equation}
where $\beta_{\mu}$ is defined in Eq.~(\ref{eq:beta}). From Eq.~(\ref{eq:F}), it
follows that the two Casimir operators ${\cal C}_1$, ${\cal C}_3$ of
Eqs.~(\ref{eq:casimir1}), (\ref{eq:casimir3}) reduce to the well-known Casimir
operators, $U = \exp({\rm i} 2\pi N)$ and ${\cal C} = F(N) - a^{\dagger} a$,
respectively (Quesne and Vansteenkiste, 1996, 1997).\par
%
%
It is worth noting that there exist other realizations of $C_{\lambda}$,
which may
be interesting in some physical applications. We shall mention here two of
them.\par
%
%
The first one uses functions of spin $s$ operators, where $s = (\lambda-1)/2$.
Denoting as usual the spin operators (generating an su(2) Lie algebra) by $S_i$,
$i=1$, 2, 3, it is obvious that the operators
\begin{equation}
  P_{\mu} = \prod_{\scriptstyle \sigma=-(\lambda-1)/2 \atop \scriptstyle
  \sigma\ne (\lambda-2\mu-1)/2}^{(\lambda-1)/2} \frac{S_3 - \sigma}
  {\frac{1}{2}(\lambda-2\mu-1) - \sigma}, \qquad \mu = 0, 1, \ldots, \lambda-1,
\end{equation}
acting in spin space, project on the spin components $\sigma = (\lambda-1)/2$,
$(\lambda-3)/2$, $\ldots$, $(\lambda-2\mu-1)/2$, $\ldots$, $-(\lambda-1)/2$,
respectively. The corresponding realization of the $C_{\lambda}$ generator
$T$ is
obtained from Eq.~(\ref{eq:T}) in the form
\begin{equation}
  T = \sum_{\mu=0}^{\lambda-1} e^{{\rm i} 2\pi \mu/\lambda} \left(
  \prod_{\scriptstyle \nu=0 \atop \scriptstyle \nu\ne\mu}^{\lambda-1}
  \frac{2S_3 - \lambda + 2\nu +1}{2(\nu - \mu)}\right).
\end{equation}
\par
%
%
By using the $(2s+1) \times (2s+1)$ matrix representation of $S_3$, $S_3 = {\rm
diag} (s, s-1, \ldots, -s)$, we get another realization of $C_{\lambda}$ in
terms of
$\lambda \times \lambda$ matrices,
\begin{equation}
  T = \sum_{\mu=0}^{\lambda-1} e^{{\rm i} 2\pi \mu/\lambda} e_{\mu+1,\mu+1},
  \qquad P_{\mu} = e_{\mu+1,\mu+1},  \label{eq:T-matrix}
\end{equation}
where $e_{ij}$ denotes the $\lambda \times \lambda$ matrix with 1 in row $i$
and column $j$, and zeros everywhere else.\par
%
%
Note that when considering such realizations of $C_{\lambda}$, the remaining
${\cal A}^{(\lambda)}_{\alpha_0 \alpha_1 \ldots \alpha_{\lambda-2}}$ generators
would either act in both configuration and spin spaces, or be $\lambda \times
\lambda$ operator-valued matrices.\par
%
%
{}For $\lambda=2$, the last relation in Eq.~(\ref{eq:alg-def1-1}) and those in
Eq.~(\ref{eq:alg-def1-2}) become
\begin{equation}
  T^2 = I, \qquad \left\{a^{\dagger}, T\right\} = 0, \qquad \left[a,
a^{\dagger} \right]
  = I + \kappa_1 T = I + \alpha_0 (P_0 - P_1),  \label{eq:lambda2}
\end{equation}
where $P_0 = (I + T)/2$, $P_1 = (I - T)/2$, and $\kappa_1$, $\alpha_0 \in \R$.
In the corresponding GDOA, the operator $T$ is given by $T = \exp({\rm i}
\pi N)$,
which amounts to Klein operator $K = (-1)^N$, since as shown in the next
section,
the eigenvalues of $N$ are integer in the ${\cal A}^{(2)}(G(N))$ unirreps.
In the
matrix realization~(\ref{eq:T-matrix}), T is represented by the Pauli spin
matrix
$\sigma_3$, while $a^{\dagger}$, $a$ can be expressed in terms of $\sigma_1$,
$\sigma_2$, and some differential operators (Bagchi, 1994).\par
%
%
{}For $\lambda=3$, the counterpart of Eq.~(\ref{eq:lambda2}) reads
\begin{equation}
  T^3 = I, \qquad a^{\dagger} T = e^{-{\rm i} 2\pi/3} T a^{\dagger},
\end{equation}
\begin{equation}
  \left[a, a^{\dagger} \right] = I + \kappa_1 T + \kappa_1^* T^2 = I +
\alpha_0 P_0
  + \alpha_1 P_1 - (\alpha_0 + \alpha_1) P_2,
\end{equation}
where  $P_0 = (I + T + T^2)/3$, $P_1 = \left(I + e^{-{\rm i} 2\pi/3} T +
e^{-{\rm i}
4\pi/3} T^2\right)/3$,
\linebreak
$P_2 = \left(I + e^{-{\rm i} 4\pi/3} T + e^{-{\rm i}
2\pi/3}
T^2\right)/3$, $\kappa_1 \in \C$, and $\alpha_0$, $\alpha_1 \in \R$. In
the GDOA realization, the operator $T$ is given by $T = \exp({\rm i} 2\pi
N/3)$, so
that $G(N) = I + 2 (\Re e\, \kappa_1) \cos(2\pi N/3) - 2 (\Im m\, \kappa_1)
\sin(2\pi N/3)$. In the matrix realization~(\ref{eq:T-matrix}), $T$ is
represented
by the matrix
${\rm diag}\left(1, e^{{\rm i} 2\pi/3}, e^{{\rm i} 4\pi/3}\right)$. Explicit
expressions of $a^{\dagger}$, $a$ are still unknown.\par
%
%
In the remainder of this paper, we shall concentrate on the abstract
definition of
${\cal A}^{(\lambda)}_{\alpha_0 \alpha_1 \ldots \alpha_{\lambda-2}}$, or
its GDOA
realization ${\cal A}^{(\lambda)}(G(N))$.\par
%
%
\section{\boldmath  UNIRREPS OF $C_{\lambda}$-EXTENDED OSCILLATOR
ALGEBRAS}
\label{sec:irreps}
\setcounter{equation}{0}
The purpose of the present section is to provide a classification of the ${\cal
A}^{(\lambda)}_{\alpha_0 \alpha_1 \ldots \alpha_{\lambda-2}}$ unirreps. To carry
out this program, it proves convenient to first consider the corresponding GDOA
${\cal A}^{(\lambda)}(G(N))$, defined in
Eqs.~(\ref{eq:T-GDOA})--(\ref{eq:G}).\par
%
%
\subsection{\boldmath Unirreps of ${\cal A}^{(\lambda)}(G(N))$}
\label{sec:irrepsA}
As a consequence of Eq.~(\ref{eq:casimir-rel}), and of the assumption
${\cal C}_2 =
I$, the first Casimir operator $U = {\cal C}_1$ of ${\cal A}^{(\lambda)}(G(N))$
reduces to $I$; hence the eigenvalues of $N$ are integer. As usual, we
shall restrict
ourselves to those unirreps wherein they are
nondegenerate (Rideau, 1992; Quesne and Vansteenkiste, 1996, 1997).$^2$\par
%
%
Let us start with a normalized simultaneous eigenvector $|c, n_0\rangle$ of the
Casimir operator ${\cal C} = {\cal C}_3$, defined in
Eq.~(\ref{eq:casimir3}), and of
the number operator $N$, corresponding to the eigenvalues $c \in \R$
and $n_0
\in \Z$, respectively. From Eqs.~(\ref{eq:alg-def2-1})--(\ref{eq:alg-def2-3}),
it results that as long as
they are nonvanishing, the vectors
\begin{equation}
  |c, n_0 + n) = \left\{\begin{array}{ll}
        \left(a^{\dagger}\right)^n |c, n_0\rangle, & \mbox{if
$n=0,1,\ldots$,} \\
        a^{-n} |c, n_0\rangle, & \mbox{if $n=-1,-2,\ldots$,}
                                 \end{array}\right.    \label{eq:unnormalized}
\end{equation}
satisfy the relations
\begin{equation}
  {\cal C} |c, n_0 + n) = c |c, n_0 + n), \qquad N |c, n_0 + n) = (n_0 + n)
|c, n_0 + n), \label{eq:unnormalized-rel-1}
\end{equation}
\begin{equation}
  a^{\dagger} a |c, n_0 + n) = \lambda_n |c, n_0 + n), \qquad a a^{\dagger}
|c, n_0 + n)
  = \lambda_{n+1} |c, n_0 + n), \label{eq:unnormalized-rel-2}
\end{equation}
where
\begin{equation}
  \lambda_n = F(n_0+n) - c.
\end{equation}
\par
%
%
In any unirrep, only nonnegative values of $\lambda_n$ are allowed. From
Eq.~(\ref{eq:F}), it is clear that the unirrep carrier space $\cal S$ is
$\Z_{\lambda}$-graded: ${\cal S} = \sum_{\mu=0}^{\lambda-1} \oplus {\cal
S}_{\mu}$, where ${\cal S}_{\mu} = \{\, |c, n_0+n) \mid n_0+n = \mu\, {\rm
mod}\,
\lambda\,\}$. Hence, we have to discuss the unitarity conditions $\lambda_n
\ge 0$
separately in each ${\cal S}_{\mu}$ subspace. Since the structure function
$F(N)$ is
an increasing linear function of $N$ in each ${\cal S}_{\mu}$, it is
obvious that the
algebra has no infinite-dimensional bounded from above (BFA) nor unbounded (UB)
unirreps (Quesne and Vansteenkiste, 1996, 1997). It therefore only remains to
successively consider the
cases of infinite-dimensional bounded from below (BFB) unirreps and of
finite-dimensional (FD) ones.\par
%
%
In the case of BFB unirreps, the eigenvalues of N are $n_0$, $n_0+1$,
$n_0+2$,~\ldots, and the unitarity conditions reduce to
\begin{equation}
  \lambda_0 = 0, \qquad \lambda_n > 0 \qquad {\rm if\ } n = 1, 2, \ldots,
\lambda-1.
  \label{eq: BFB-unitarity}
\end{equation}
The first condition in Eq.~(\ref{eq: BFB-unitarity}) fixes the Casimir operator
eigenvalue,
\begin{equation}
  c = n_0 + \beta_{\mu_0},  \label{eq:BFB-c}
\end{equation}
where $\mu_0 \in \{0, 1, \ldots, \lambda-1\}$ is defined by
\begin{equation}
  n_0 = \mu_0\, {\rm mod}\, \lambda,  \label{eq:muzero}
\end{equation}
while the second condition yields some restrictions on the algebra parameters,
\begin{equation}
  \overline{\beta}_{\nu} - \overline{\beta}_{\mu_0} + 1 > 0, \qquad {\rm
if\ } \nu =
  0, 1, \ldots, \mu_0 - 1, \label{eq:BFB-beta-1}
\end{equation}
\begin{equation}
  \overline{\beta}_{\nu} - \overline{\beta}_{\mu_0} > 0, \qquad {\rm if\ } \nu =
  \mu_0 + 1, \mu_0 + 2, \ldots, \lambda - 1, \label{eq:BFB-beta-2}
\end{equation}
where
\begin{equation}
  \overline{\beta}_{\mu} = \frac{\beta_{\mu} + \mu}{\lambda}.
\label{eq:betabar}
\end{equation}
In terms of the $\alpha_{\mu}$'s, Eqs.~(\ref{eq:BFB-c}), (\ref{eq:BFB-beta-1}),
and~(\ref{eq:BFB-beta-2}) can
be rewritten as
\begin{equation}
  c = n_0 + \sum_{\nu=0}^{\mu_0-1} \alpha_{\nu},
\end{equation}
and
\begin{equation}
  \alpha_{\nu} < \lambda - \mu_0 + \nu - \sum_{\rho=\nu+1}^{\mu_0-1}
  \alpha_{\rho}, \qquad {\rm if\ } \nu =
  0, 1, \ldots, \mu_0 - 1,
\end{equation}
\begin{equation}
  \alpha_{\nu} > \mu_0 - \nu - 1 - \sum_{\rho=\mu_0}^{\nu-1} \alpha_{\rho},
  \qquad {\rm if\ } \nu = \mu_0, \mu_0 + 1, \ldots, \lambda - 2,
\end{equation}
respectively.\par
%
%
Whenever the unitarity conditions are satisfied, normalized basis states of
$\cal
S$ can be constructed from the vectors~(\ref{eq:unnormalized}), and are given by
\begin{equation}
  |c, n_0+n\rangle = \left[{\cal N}_n(c, n_0)\right]^{-1/2} |c, n_0+n),
\qquad n = 0, 1,
  2, \ldots,    \label{eq:normalized}
\end{equation}
where the normalization coefficient is
\begin{equation}
  {\cal N}_n(c, n_0) = \prod_{i=1}^n \lambda_i = \prod_{i=1}^n [F(n_0+i) - c].
  \label{eq:normalization}
\end{equation}
By writing $n$ as $n = k\lambda + \mu$, where $\mu \in \{0, 1, \ldots,
\lambda-1\}$, and $k \in \N$, ${\cal N}_n(c, n_0)$ can be expressed
in terms
of gamma functions as
\begin{eqnarray}
  {\cal N}_{k\lambda+\mu}(c, n_0) & = & \lambda^{k\lambda+\mu} \left(
       \prod_{\nu=0}^{\mu_0+\mu} \Gamma(\overline{\beta}_{\nu} -
       \overline{\beta}_{\mu_0} + k + 1)\right)
       \left(\prod_{\nu=\mu_0+\mu+1}^{\lambda-1}
       \Gamma(\overline{\beta}_{\nu} - \overline{\beta}_{\mu_0} + k )\right)
       \nonumber\\
  & & \mbox{} \times \left(\prod_{\nu=0}^{\mu_0}
       \Gamma(\overline{\beta}_{\nu} - \overline{\beta}_{\mu_0} + 1)\right)^{-1}
       \left(\prod_{\nu=\mu_0+1}^{\lambda-1}
       \Gamma(\overline{\beta}_{\nu} - \overline{\beta}_{\mu_0})\right)^{-1},
       \nonumber\\
  & &{\rm if\ } \mu = 0, 1, \ldots, \lambda - \mu_0 - 1, \nonumber\\
  & = & \lambda^{k\lambda+\mu} \left( \prod_{\nu=0}^{\mu_0+\mu-\lambda}
       \Gamma(\overline{\beta}_{\nu} - \overline{\beta}_{\mu_0} + k + 2)\right)
       \left(\prod_{\nu=\mu_0+\mu-\lambda+1}^{\lambda-1}
       \Gamma(\overline{\beta}_{\nu} - \overline{\beta}_{\mu_0} + k + 1 )\right)
       \nonumber\\
  & & \mbox{} \times \left(\prod_{\nu=0}^{\mu_0}
       \Gamma(\overline{\beta}_{\nu} - \overline{\beta}_{\mu_0} + 1)\right)^{-1}
       \left(\prod_{\nu=\mu_0+1}^{\lambda-1}
       \Gamma(\overline{\beta}_{\nu} - \overline{\beta}_{\mu_0})\right)^{-1},
       \nonumber\\
  & &{\rm if\ } \mu = \lambda - \mu_0, \lambda - \mu_0 + 1, \ldots, \lambda
       - 1.
\end{eqnarray}
\par
%
%
In the case of FD unirreps, the eigenvalues of N are $n_0$, $n_0+1$, \ldots,
$n_0+d-1$, where the dimension $d$ may only take values in the set $\{1, 2,
\ldots,
\lambda-1\}$. The unitarity conditions are then given by
\begin{equation}
  \lambda_0 = 0, \qquad \lambda_n > 0 \qquad {\rm if\ } n = 1, 2, \ldots,
d-1, \qquad
  \lambda_d = 0.
\end{equation}
Defining $\mu_0$ and $\overline{\beta}_{\mu}$ as before by
Eqs.~(\ref{eq:muzero})
and (\ref{eq:betabar}), respectively, we obtain
\begin{equation}
  c = n_0 + \beta_{\mu_0}, \label{eq:FD-cond1-1}
\end{equation}
\begin{equation}
  \overline{\beta}_{\nu} - \overline{\beta}_{\mu_0} > 0, \qquad {\rm if\ } \nu =
  \mu_0 + 1, \mu_0 + 2, \ldots, \mu_0 + d - 1, \label{eq:FD-cond1-2}
\end{equation}
\begin{equation}
  \overline{\beta}_{\mu_0+d} - \overline{\beta}_{\mu_0} = 0,
  \label{eq:FD-cond1-3}
\end{equation}
for $\mu_0 = 0$, 1, \ldots,~$\lambda - d - 1$, and
\begin{equation}
  c = n_0 + d + \beta_{\mu_0-\lambda+d}, \label{eq:FD-cond2-1}
\end{equation}
\begin{equation}
  \overline{\beta}_{\nu} - \overline{\beta}_{\mu_0-\lambda+d} > 0, \qquad
  {\rm if\ } \nu = 0, 1, \ldots, \mu_0 - \lambda + d - 1, \label{eq:FD-cond2-2}
\end{equation}
\begin{equation}
  \overline{\beta}_{\nu} - \overline{\beta}_{\mu_0} > 0, \qquad {\rm if\ } \nu =
  \mu_0 + 1, \mu_0 + 2, \ldots, \lambda - 1, \label{eq:FD-cond2-3}
\end{equation}
\begin{equation}
  \overline{\beta}_{\mu_0-\lambda+d} - \overline{\beta}_{\mu_0} + 1 = 0,
  \label{eq:FD-cond2-4}
\end{equation}
for $\mu_0 = \lambda - d$, $\lambda - d + 1$, \ldots,~$\lambda - 1$. In terms of
the algebra parameters $\alpha_{\mu}$,
Eqs.~(\ref{eq:FD-cond1-1})--(\ref{eq:FD-cond1-3}), and
Eqs.~(\ref{eq:FD-cond2-1})--(\ref{eq:FD-cond2-4}) become
\begin{equation}
  c = n_0 + \sum_{\nu=0}^{\mu_0-1} \alpha_{\nu},
\end{equation}
\begin{equation}
  \alpha_{\nu} > \mu_0 - \nu - 1 - \sum_{\rho=\mu_0}^{\nu-1} \alpha_{\rho},
\qquad
  {\rm if\ } \nu = \mu_0, \mu_0 + 1, \ldots, \mu_0 + d - 2,
\end{equation}
\begin{equation}
  \alpha_{\mu_0+d-1} = - d - \sum_{\rho=\mu_0}^{\mu_0+d-2} \alpha_{\rho},
\end{equation}
for $\mu_0 = 0$, 1, \ldots,~$\lambda - d - 1$, and
\begin{equation}
  c = n_0 + d + \sum_{\nu=0}^{\mu_0-\lambda+d-1} \alpha_{\nu},
\end{equation}
\begin{equation}
  \alpha_{\nu} < \lambda - \mu_0 + \nu - d - \sum_{\rho=\nu+1}^{\mu_0-\lambda
  +d-1} \alpha_{\rho}, \qquad {\rm if\ } \nu = 0, 1, \ldots, \mu_0 -
\lambda + d - 1,
\end{equation}
\begin{equation}
  \alpha_{\nu} > \mu_0 - \nu - 1 - \sum_{\rho=\mu_0}^{\nu-1} \alpha_{\rho},
\qquad
  {\rm if\ } \nu = \mu_0, \mu_0 + 1, \ldots, \lambda - 2,
\end{equation}
\begin{equation}
  \alpha_{\mu_0-1} = d - \sum_{\rho=\mu_0-\lambda+d}^{\mu_0-2} \alpha_{\rho},
\end{equation}
for $\mu_0 = \lambda - d$, $\lambda - d + 1$, \ldots,~$\lambda - 1$,
respectively.\par
%
%
Normalized basis states of the carrier space $\cal S$ of a $d$-dimensional
unirrep
are given by Eqs.~(\ref{eq:normalized}) and~(\ref{eq:normalization}), where
$n$ now
runs over the range $n=0$, 1, \ldots,~$d-1$. The corresponding normalization
coefficient ${\cal N}_n(c,n_0)$ can be rewritten as
\begin{equation}
  {\cal N}_n(c,n_0) = \lambda^n \prod_{\nu=\mu_0+1}^{\mu_0+n} \left(
  \overline{\beta}_{\nu} - \overline{\beta}_{\mu_0}\right),
\end{equation}
for $\mu_0 = 0$, 1, \ldots,~$\lambda - d - 1$, and
\begin{eqnarray}
  {\cal N}_n(c,n_0) & = & \lambda^n \prod_{\nu=\mu_0+1}^{\mu_0+n} \left(
        \overline{\beta}_{\nu} - \overline{\beta}_{\mu_0-\lambda+d} - 1\right),
        \qquad {\rm if\ } n = 1, 2, \ldots, \lambda - \mu_0 - 1, \nonumber\\
  & = & \lambda^n  \left(\prod_{\nu=0}^{\mu_0+n-\lambda} \left(
        \overline{\beta}_{\nu} -
\overline{\beta}_{\mu_0-\lambda+d}\right)\right)
        \left(\prod_{\nu=\mu_0+1}^{\lambda-1} \left(\overline{\beta}_{\nu} -
        \overline{\beta}_{\mu_0-\lambda+d} - 1\right)\right), \nonumber\\
  & & {\rm if\ } n = \lambda - \mu_0, \lambda - \mu_0 + 1, \ldots, d - 1,
\end{eqnarray}
for $\mu_0 = \lambda - d$, $\lambda - d + 1$, \ldots,~$\lambda - 1$.\par
%
%
In Tables~I,~II, and~III, the detailed unirrep
classification is given for $\lambda=2$, $\lambda=3$, and $\lambda=4$,
respectively.\par
%
%
Of special interest in physical applications are the Fock-space unirreps,
characterized by $c = n_0 = 0$. Since in this case $\mu_0 = 0$, such
representations exist whenever the algebra parameters satisfy the conditions
\begin{equation}
  \sum_{\rho=0}^{\nu} \alpha_{\rho} > - \nu - 1, \qquad {\rm if\ }\nu = 0,
1, \ldots,
  \lambda - 2,  \label{eq:Fock-cond}
\end{equation}
in the BFB case, and
\begin{equation}
  \sum_{\rho=0}^{\nu} \alpha_{\rho} > - \nu - 1, \qquad {\rm if\ }\nu = 0,
1, \ldots,
  d - 2,
\end{equation}
\begin{equation}
  \sum_{\rho=0}^{d-1} \alpha_{\rho} = - d,
\end{equation}
in the FD one. The former are of bosonic type. Apart from the trivial
one-dimensional unirrep, the latter are of fermionic or order-$p$-parafermionic
type, according to whether $d=2$ or $d=p+1 \ge 3$. Note that parafermionic-type
unirreps only appear for $\lambda \ge 4$.\par
%
%
In the bosonic Fock-space representation, it may be interesting to consider
a bosonic
oscillator Hamiltonian (Quesne and Vansteenkiste, 1998), defined in appropriate
units by
\begin{equation}
  H_0 = \frac{1}{2} \left\{a, a^{\dagger}\right\}.   \label{eq:Hzero}
\end{equation}
By using
Eqs.~(\ref{eq:alg-def2-1})--(\ref{eq:alg-def2-3}),
and~(\ref{eq:casimir3}), $H_0$ can
be rewritten in the equivalent forms
\begin{equation}
  H_0 = a^{\dagger} a + \frac{1}{2} \left(I + \sum_{\mu=0}^{\lambda-1}
\alpha_{\mu}
  P_{\mu}\right) = N + \frac{1}{2} I + \sum_{\mu=0}^{\lambda-1} \gamma_{\mu}
  P_{\mu},  \label{eq:Hzerobis}
\end{equation}
where the parameters~$\gamma_{\mu}$ are defined by
\begin{equation}
  \gamma_{\mu} \equiv \frac{1}{2} (\beta_{\mu} + \beta_{\mu+1}) =
  \left\{\begin{array}{ll}
        \frac{1}{2} \alpha_0, & \mbox{if $\mu=0$}, \\
        \sum_{\nu=0}^{\mu-1} \alpha_{\nu} + \frac{1}{2} \alpha_{\mu}, & \mbox{if
              $\mu=1, 2, \ldots, \lambda-1$}.
  \end{array}\right.   \label{eq:gamma}
\end{equation}
The latter satisfy the relation
\begin{equation}
  \sum_{\mu=0}^{\lambda-1}\, (-1)^{\mu} \gamma_{\mu} = 0,
\end{equation}
deriving from Eq.~(\ref{eq:alpha-cond}), as well as the inequalities
\begin{equation}
  \gamma_{\mu} > - \frac{1}{2} (2\mu + 1), \qquad {\rm if\ } \mu = 0, 1, \ldots,
  \lambda - 2, \label{eq:Fock-condbis-1}
\end{equation}
\begin{equation}
  \gamma_{\lambda-1} > - \frac{1}{2} (\lambda - 1), \label{eq:Fock-condbis-2}
\end{equation}
coming from conditions~(\ref{eq:Fock-cond}).\par
%
%
The states $|n\rangle = |k\lambda + \mu\rangle$, given by
Eq.~(\ref{eq:normalized})
where $c = n_0 = 0$, are the eigenstates of $H_0$, corresponding to the
eigenvalues
\begin{equation}
  E_{k\lambda+\mu} = k\lambda + \mu + \gamma_{\mu} + \frac{1}{2}, \qquad k = 0,
 1, 2, \ldots, \qquad \mu = 0, 1, \ldots, \lambda-1.
\end{equation}
In each ${\cal F}_{\mu} = \{\, |k\lambda+\mu\rangle \mid k=0, 1, 2, \ldots \,\}$
subspace of the $\Z_{\lambda}$-graded Fock space~${\cal F} =
\sum_{\mu=0}^{\lambda-1} \oplus {\cal F}_{\mu}$, the spectrum of~$H_0$ is
harmonic, but the $\lambda$ infinite sets of equally spaced energy levels,
corresponding to $\mu=0$, 1, $\ldots$,~$\lambda-1$, may be shifted with respect
to each other by some amounts depending upon the algebra parameters $\alpha_0$,
$\alpha_1$, $\ldots$,~$\alpha_{\lambda-2}$, through their linear combinations
$\gamma_0$, $\gamma_1$, $\ldots$,~$\gamma_{\lambda-1}$. As a result, one may
get nondegenerate spectra, as well as spectra exhibiting some $(\nu+1)$-fold
degeneracies, where $\nu$ may take any value in the set $\{1, 2, \ldots,
\lambda-1\}$ (Quesne and Vansteenkiste, 1998, 1999).\par
%
%
\subsection{\boldmath Unirreps of ${\cal A}^{(\lambda)}_{\alpha_0 \alpha_1
\ldots
\alpha_{\lambda-2}}$}
\label{sec:irrepsB}
Let us now turn ourselves to the general case of ${\cal A}^{(\lambda)}_{\alpha_0
\alpha_1 \ldots \alpha_{\lambda-2}}$, defined in Eqs.~(\ref{eq:alg-def1-1})
and~(\ref{eq:alg-def1-2}).
Since we
do not assume ${\cal C}_2 = I$, the eigenvalues of $N$ are not restricted
to integer
values anymore. It can however be shown that they are discrete. The proof
proceeds
as in Jordan {\em et al.} (1963) and Quesne and Vansteenkiste (1997), and can be
summarized as
follows. The Casimir operator ${\cal C}_1$, defined in
Eq.~(\ref{eq:casimir1}), is
unitary, so that in any given unirrep its eigenvalue can be written as
$\exp({\rm i}
2\pi \nu_0)$, where $\nu_0 \in \R$. On the other hand, the eigenvalues of
${\cal C}_1$ can be determined from those of the Hermitian operator $N$. The
spectral mapping theorem leads to eigenvalues of ${\cal C}_1$ of the form
$\exp({\rm i} 2\pi x)$, where $x \in \R$ are the eigenvalues of $N$. The
equivalence of the two expressions for the eigenvalues of ${\cal C}_1$ implies
that $x = \nu_0 + n$, $n \in \Z$, in any given unirrep, which
completes the
proof. As in Section~\ref{sec:irrepsA}, we shall restrict ourselves to those
unirreps
wherein the spectrum of $N$ is not only discrete, but also nondegenerate.\par
%
%
As in Eq.~(\ref{eq:unnormalized}), the carrier space of any ${\cal
A}^{(\lambda)}_{\alpha_0 \alpha_1 \ldots \alpha_{\lambda-2}}$ unirrep can be
constructed by successive applications of $a^{\dagger}$ or $a$ on a normalized
simultaneous eigenvector $|c, \gamma, \nu_0\rangle$ of $N$ and of the Casimir
operators ${\cal C}_1$, ${\cal C}_2$, ${\cal C}_3$, defined in
Eqs.~(\ref{eq:casimir1})--(\ref{eq:casimir3}),
\begin{equation}
  N |c, \gamma, \nu_0\rangle = \nu_0 |c, \gamma, \nu_0\rangle,
\end{equation}
\begin{equation}
  {\cal C}_1 |c, \gamma, \nu_0\rangle = e^{{\rm i} 2\pi r_0} |c, \gamma,
  \nu_0\rangle,
\end{equation}
\begin{equation}
  {\cal C}_2 |c, \gamma, \nu_0\rangle = e^{{\rm i} 2\pi (- r_0
+\gamma)/\lambda} |c,
  \gamma, \nu_0\rangle,
\end{equation}
\begin{equation}
  {\cal C}_3 |c, \gamma, \nu_0\rangle = c |c, \gamma, \nu_0\rangle.
\end{equation}
Here $c$, $\nu_0 \in \R$, $\gamma \in \{0, 1, \ldots, \lambda-1\}$,
$r_0 \in
[0, 1)$ is defined by
\begin{equation}
  \nu_0 = n_0 + r_0, \qquad n_0 \in \Z,
\end{equation}
and the eigenvalue of ${\cal C}_2$ is determined from
Eq.~(\ref{eq:casimir-rel}).\par
%
%
Let us now introduce some new operators and parameters, defined by
\begin{equation}
  N' \equiv N - r_0 I, \qquad a^{\prime\dagger} \equiv a^{\dagger}, \qquad
a' \equiv a,
  \qquad T' \equiv e^{-{\rm i} 2\pi \gamma/\lambda} T,
\end{equation}
\begin{equation}
  \kappa'_{\mu} \equiv e^{{\rm i} 2\pi \mu \gamma/\lambda} \kappa_{\mu} =
  \kappa^{\prime*}_{\lambda-\mu},
\end{equation}
from which we obtain
\begin{equation}
  P'_{\mu} \equiv \frac{1}{\lambda} \sum_{\nu=0}^{\lambda-1} e^{-{\rm i}
2\pi \mu
  \nu/\lambda} T^{\prime\nu} = P_{\mu+\gamma},
\end{equation}
\begin{equation}
  \alpha'_{\mu} \equiv \sum_{\nu=1}^{\lambda-1} e^{{\rm i} 2\pi \mu \nu/\lambda}
  \kappa'_{\nu} = \alpha_{\mu+\gamma} = \alpha^{\prime*}_{\mu}.
  \label{eq:alphaprime}
\end{equation}
It is obvious that $N'$, $a^{\prime\dagger}$, $a'$, $T'$ (or $P'_{\mu}$) satisfy
the
defining relations~(\ref{eq:alg-def1-1}) and~(\ref{eq:alg-def1-2})
(or~(\ref{eq:alg-def2-1})--(\ref{eq:alg-def2-3})) of
${\cal
A}^{(\lambda)}_{\alpha'_0 \alpha'_1 \ldots \alpha'_{\lambda-2}}$, where the
primed
parameters $\alpha'_{\mu}$ are given by Eq.~(\ref{eq:alphaprime}). The
corresponding Casimir operators ${\cal C}'_1$, ${\cal C}'_2$, ${\cal C}'_3$
are found
to be expressible in terms of the old ones ${\cal C}_1$, ${\cal C}_2$,
${\cal C}_3$,
\begin{equation}
  {\cal C}'_1 \equiv e^{{\rm i} 2\pi N'} = e^{-{\rm i} 2\pi r_0} {\cal C}_1,
\end{equation}
\begin{equation}
  {\cal C}'_2 \equiv e^{-{\rm i} 2\pi N'/\lambda}\, T' = e^{{\rm i} 2\pi (r_0 -
  \gamma)/\lambda} {\cal C}_2,
\end{equation}
\begin{equation}
  {\cal C}'_3 \equiv N' + \sum_{\mu=0}^{\lambda-1} \beta'_{\mu} P'_{\mu} -
  a^{\prime\dagger} a' = {\cal C}_3 - (r_0 + \beta_{\gamma}) I,
\end{equation}
where $\beta'_{\mu} \equiv \sum_{\nu=0}^{\mu-1} \alpha'_{\nu} = \beta_{\mu +
\gamma} - \beta_{\gamma}$.\par
%
%
Hence, the simultaneous eigenvector $|c, \gamma, \nu_0\rangle$ of $N$, ${\cal
C}_1$, ${\cal C}_2$, ${\cal C}_3$ is also a simultaneous eigenvector of
$N'$, ${\cal
C}'_1$, ${\cal C}'_2$, ${\cal C}'_3$, satisfying the relations
\begin{equation}
  N' |c, \gamma, \nu_0\rangle = n_0 |c, \gamma, \nu_0\rangle,
\end{equation}
\begin{equation}
  {\cal C}'_1 |c, \gamma, \nu_0\rangle = {\cal C}'_2 |c, \gamma, \nu_0\rangle =
  |c, \gamma, \nu_0\rangle,
\end{equation}
\begin{equation}
  {\cal C}'_3 |c, \gamma, \nu_0\rangle = c' |c, \gamma, \nu_0\rangle,
\end{equation}
where
\begin{equation}
  c' = c - r_0 - \beta_{\gamma}.  \label{eq:cprime}
\end{equation}
{}From Section~\ref{sec:irrepsA}, it follows that such a state may be identified
with the
starting eigenvector $|c', n_0\rangle$ of some unirrep of the GDOA ${\cal
A}^{(\lambda)}(G'(N'))$, where $G'(N')  = I + \sum_{\mu=0}^{\lambda-1}
\alpha'_{\mu}
P'_{\mu}$. Since $a^{\prime\dagger} = a^{\dagger}$ and $a' = a$, this
correspondence
between $|c, \gamma, \nu_0\rangle$ and $|c', n_0\rangle$ extends to the
remaining
basis states of the ${\cal A}^{(\lambda)}_{\alpha_0 \alpha_1 \ldots
\alpha_{\lambda-2}}$ and ${\cal A}^{(\lambda)}(G'(N'))$ unirreps built on such
vectors, respectively.\par
%
%
We conclude that to every BFB (or FD) unirrep of ${\cal A}^{(\lambda)}(G'(N'))$,
specified by some minimal $N'$ eigenvalue $n_0 \in \Z$ (and some dimension
$d$), we may associate an infinite number of BFB (or FD) unirreps of ${\cal
A}^{(\lambda)}_{\alpha_0 \alpha_1 \ldots \alpha_{\lambda-2}}$, characterized by
minimal $N$ eigenvalues $\nu_0 = n_0 + r_0$, $r_0 \in [0, 1)$, as well as ${\cal
C}_2$ eigenvalues $\exp[{\rm i} 2\pi(- r_0 + \gamma)/\lambda]$, $\gamma \in \{0,
1, \ldots, \lambda-1\}$ (and the same dimension $d$). The eigenvalues of the
corresponding Casimir operators ${\cal C}'_3 = {\cal C}'$ and ${\cal C}_3$ are
connected by Eq.~(\ref{eq:cprime}). Furthermore, all the ${\cal
A}^{(\lambda)}_{\alpha_0 \alpha_1 \ldots \alpha_{\lambda-2}}$ unirreps are
obtained by this mapping procedure.\par
%
%
\section{\boldmath APPLICATION OF $C_{\lambda}$-EXTENDED OSCILLATOR
ALGEBRAS TO PSSQM OF ORDER $\lowercase{p} = \lambda - 1$}
\label{sec:para}
\setcounter{equation}{0}
PSSQM of order two was introduced by Rubakov and Spiridonov (1988) as a
generalization of SSQM (Witten, 1981), obtained by combining standard
fermions with
parafermions of order two (Green, 1953; Ohnuki and Kamefuchi, 1982) instead of
standard fermions. Its
extension
to arbitrary order $p$, due to Khare (1992, 1993), is described in terms of
parasupercharge operators $Q$, $Q^{\dagger}$, and a parasupersymmetric
Hamiltonian $\cal H$, satisfying the relations
\begin{equation}
  Q^{p+1} = 0 \qquad ({\rm with\ } Q^p \ne 0),  \label{eq:Q-power}
\end{equation}
\begin{equation}
  [{\cal H}, Q] = 0,  \label{eq:H-Q}
\end{equation}
\begin{equation}
  Q^p Q^{\dagger} + Q^{p-1} Q^{\dagger} Q + \cdots + Q Q^{\dagger} Q^{p-1} +
  Q^{\dagger} Q^p = 2p Q^{p-1} {\cal H},  \label{eq:multi}
\end{equation}
and their Hermitian conjugates.\par
%
%
As shown by Bagchi {\em et al.} (1997), PSSQM of order $p$ can be
reformulated in terms of $p$ super (rather than parasuper) charges
$Q_{\nu}$, $\nu
= 1$, 2, \ldots,~$p$, all of which satisfy $Q_{\nu}^2 = 0$ and commute with
${\cal
H}$. However, unlike in usual SSQM, $\cal H$ cannot be simply expressed in
terms of
the $p$ supercharges (except in a very special case to be reviewed below). More
specifically, let us set
\begin{equation}
  Q = \sum_{\nu=1}^{p} \sigma_{\nu} Q_{\nu},  \label{eq:Q-bagchi}
\end{equation}
where $\sigma_{\nu}$ are some complex constants, and $Q_{\nu}$, $\nu = 1$, 2,
\ldots,~$p$, are assumed to satisfy the relations
\begin{equation}
  Q_{\nu} Q_{\nu'} = \delta_{\nu',\nu+1} Q_{\nu} Q_{\nu+1},
  \label{eq:super-cond-1}
\end{equation}
\begin{equation}
  Q_{\nu} Q_{\nu'}^{\dagger} = \delta_{\nu',\nu} Q_{\nu} Q_{\nu}^{\dagger},
  \label{eq:super-cond-2}
\end{equation}
\begin{equation}
  Q_{\nu}^{\dagger} Q_{\nu'} = \delta_{\nu',\nu} Q_{\nu}^{\dagger} Q_{\nu}.
  \label{eq:super-cond-3}
\end{equation}
Then, the operator $Q$, defined in Eq.~(\ref{eq:Q-bagchi}), satisfies
Eqs.~(\ref{eq:Q-power})--(\ref{eq:multi}) if
\begin{equation}
  \sigma_{\nu} \ne 0, \qquad \nu = 1, 2, \ldots, p, \label{eq:super-condbis1}
\end{equation}
\begin{equation}
  [{\cal H}, Q_{\nu}] = 0,  \qquad \nu = 1, 2, \ldots, p,
\label{eq:super-condbis2}
\end{equation}
\begin{equation}
  \left(\prod_{\nu=1}^{p-1} \sigma_{\nu}\right) Q_1 \Sigma +
  \left(\prod_{\nu=2}^{p} \sigma_{\nu}\right) \Sigma Q_p = 2p \left[
  \left(\prod_{\nu=1}^{p-1} \sigma_{\nu}\right) Q_1 Q_2 \ldots Q_{p-1} +
  \left(\prod_{\nu=2}^{p} \sigma_{\nu}\right) Q_2 Q_3 \ldots Q_p\right]
{\cal H},
  \label{eq:super-condbis3}
\end{equation}
where
\begin{equation}
  \Sigma \equiv |\sigma_1|^2 Q_1^{\dagger} Q_1 Q_2 \ldots Q_{p-1} +
  \sum_{\nu=2}^{p-1} |\sigma_{\nu}|^2 Q_2 Q_3 \ldots Q_{\nu} Q_{\nu}^{\dagger}
  Q_{\nu} Q_{\nu+1} \ldots Q_{p-1} + |\sigma_p|^2 Q_2 Q_3 \ldots Q_p
  Q_p^{\dagger}.
\end{equation}
\par
%
%
In the standard realization of PSSQM related to parafermions of order
$p$ (Khare, 1992, 1993), $\sigma_{\nu} = 1$, and $Q_{\nu}$,
$Q_{\nu}^{\dagger}$, $\cal
H$ are represented by $(p+1) \times (p+1)$ matrices, whose elements are
\begin{equation}
  (Q_{\nu})_{\alpha,\beta} = (P - {\rm i} W_{\beta}) \delta_{\alpha,\beta+1}
  \delta_{\beta,p+1-\nu},  \label{eq:standard1}
\end{equation}
\begin{equation}
  \left(Q_{\nu}^{\dagger}\right)_{\alpha,\beta} = (P + {\rm i} W_{\alpha})
  \delta_{\alpha,p+1-\nu} \delta_{\beta,\alpha+1},    \label{eq:standard2}
\end{equation}
\begin{equation}
  ({\cal H})_{\alpha,\beta} = {\cal H}_{\alpha} \delta_{\alpha,\beta},
  \label{eq:standard3}
\end{equation}
where $\alpha$, $\beta = 1$, 2, \ldots,~$p+1$. Here $P = - {\rm i}
\partial/\partial
x$ is the momentum operator, $W_{\nu}(x)$, $\nu = 1$, 2, \ldots,~$p$, are
superpotentials, and
\begin{equation}
  {\cal H}_{\nu} = \frac{1}{2} \left(P^2 + W_{\nu}^2 - W_{\nu}' +
C_{\nu}\right),
  \qquad \nu = 1, 2, \ldots, p, \label{eq:standard-PSSQM2-1}
\end{equation}
\begin{equation}
  {\cal H}_{p+1} = \frac{1}{2} \left(P^2 + W_p^2 + W_p' + C_p\right),
  \label{eq:standard-PSSQM2-2}
\end{equation}
with $C_{\nu} \in \R$. The operator-valued matrices~(\ref{eq:standard1})
and~(\ref{eq:standard2}) automatically satisfy
Eqs.~(\ref{eq:super-cond-1})--(\ref{eq:super-cond-3}), while
Eqs.~(\ref{eq:super-condbis2}) and~(\ref{eq:super-condbis3}) impose the
conditions
\begin{equation}
  W_{\nu}^2 + W_{\nu}' + C_{\nu} = W_{\nu+1}^2 - W_{\nu+1}' + C_{\nu+1}, \qquad
  \nu = 1, 2, \ldots, p-1,  \label{eq:potential-cond}
\end{equation}
and
\begin{equation}
  \sum_{\nu=1}^{p} C_{\nu} = 0,  \label{eq:constant-cond}
\end{equation}
respectively.\par
%
%
{}For arbitrary $W_{\nu}$'s satisfying Eqs.~(\ref{eq:potential-cond})
and~(\ref{eq:constant-cond}), the spectrum of the parasupersymmetric Hamiltonian
$\cal H$ is ($p+1$)-fold degenerate at least starting from the $p$th
excited state
onwards. The nature of the ground and the first ($p-1$) excited states however
depends on the specific form of the
$W_{\nu}$'s. For the special choice $W_1 = W_2 = \cdots = W_p = \omega x$, $\cal
H$ becomes the parasupersymmetric oscillator Hamiltonian, which can be realized
in terms of bosons and parafermions of order $p$. Its ground state is
nondegenerate, and has a negative energy, while the $\nu$th excited state
for $\nu
= 1$, 2, \ldots,~$p-1$, is ($\nu+1$)-fold degenerate.\par
%
%
We now plan to show that the PSSQM algebra~(\ref{eq:Q-power})--(\ref{eq:multi})
can be realized in terms of the generators of ${\cal A}^{(\lambda)}(G(N))$,
$\lambda
= p+1$, in their bosonic Fock-space representation (thence the parameters
$\alpha_0$, $\alpha_1$, \ldots,~$\alpha_{\lambda-2}$ satisfy
Eq.~(\ref{eq:Fock-cond})). This will prove that PSSQM of arbitrary order
$p$ can be
bosonized, as is the case for standard SSQM (Brzezi\'nski {\em et al.},
1993; Plyushchay, 1996a, b; Beckers {\em et al.}, 1997), and PSSQM
of order two (Quesne and Vansteenkiste, 1998).\par
%
%
In view of the results previously obtained for $p=2$ (Quesne and Vansteenkiste,
1998), let
us take
as ans\"atze for the operators $Q$ and $\cal H$ the expressions
\begin{equation}
  Q = \sum_{\nu=1}^{p} \eta_{\mu+\nu}\, a^{\dagger} P_{\mu+\nu},
\label{eq:Q-ansatz}
\end{equation}
\begin{equation}
  {\cal H} = H_0 + \frac{1}{2} \sum_{\nu=0}^p r_{\nu} P_{\nu},
\label{eq:H-ansatz}
\end{equation}
where $H_0$ is the bosonic oscillator Hamiltonian~(\ref{eq:Hzero}) associated
with the algebra ${\cal A}^{(p+1)}(G(N))$, $\eta_{\mu+\nu}$, $\nu = 1$, 2,
\ldots,~$p$, are some complex constants, and $r_{\nu}$, $\nu = 0$, 1,
\ldots, $p$,
some real ones. The purpose of the last term on the right-hand side of
Eq.~(\ref{eq:H-ansatz}) is to make the $p+1$ families of $H_0$ equally spaced
eigenvalues coincide at least starting from the $p$th excited state
onwards. Note
that in Eqs.~(\ref{eq:Q-ansatz}) and~(\ref{eq:H-ansatz}), $\mu$ takes some
fixed, arbitrary value in
the set
$\{0, 1, \ldots, p\}$. The operators $Q$, $Q^{\dagger}$, $\cal H$, and all the
quantities to be considered hereafter, depend on this $\mu$ value, although for
simplicity's sake we chose not to explicitly exhibit such a dependence by
appending
a $\mu$ index to them.\par
%
%
It is straightforward to see that the operators
\begin{equation}
  Q_{\nu} = a^{\dagger} P_{p+1+\mu-\nu}, \qquad \nu = 1, 2, \ldots, p,
\label{eq:supercharge}
\end{equation}
satisfy Eqs.~(\ref{eq:super-cond-1})--(\ref{eq:super-cond-3}); hence $Q$, as
defined by Eq.~(\ref{eq:Q-ansatz}), can be written in the
form~(\ref{eq:Q-bagchi}) by
setting
\begin{equation}
  \sigma_{\nu} = \eta_{p+1+\mu-\nu}, \qquad \nu = 1, 2, \ldots, p.
\end{equation}
Equation~(\ref{eq:super-condbis1}) leads to the restriction
\begin{equation}
  \eta_{\mu+\nu} \ne 0, \qquad \nu = 1, 2, \ldots, p.
\end{equation}
After some calculations, one finds that Eqs.~(\ref{eq:super-condbis2})
and~(\ref{eq:super-condbis3}) are equivalent to the conditions
\begin{equation}
  r_{\mu+\nu} = 2 + \alpha_{\mu+\nu} + \alpha_{\mu+\nu+1} + r_{\mu+\nu+1},
\qquad \nu = 1,
  2, \ldots, p,  \label{eq:r-cond}
\end{equation}
and
\begin{equation}
  \sum_{\nu=1}^p |\eta_{\mu+\nu}|^2 = 2p, \label{eq:eta-cond1}
\end{equation}
\begin{equation}
  \sum_{\nu=2}^p |\eta_{\mu+\nu}|^2 \left(\nu - 1 + \sum_{\rho=0}^{\nu-2}
  \alpha_{\mu+\rho+2} \right) = p (1 + \alpha_{\mu+2} + r_{\mu+2}),
  \label{eq:eta-cond2}
\end{equation}
respectively.\par
%
%
Equation~(\ref{eq:r-cond}) is a nonhomogeneous system of $p$ linear equations in
($p+1$) unknowns $r_{\mu+\nu}$, $\nu = 0$, 1, \ldots,~$p$. Its solution
yields $p$
of them in terms of the remaining one, e.g., $r_{\mu}$, $r_{\mu+1}$,
$r_{\mu+3}$,
\ldots, $r_{\mu+p}$ in terms of $r_{\mu+2}$:
\begin{eqnarray}
  r_{\mu} & = & - 2(p-1) - \alpha_{\mu} - \alpha_{\mu+2} - 2 \sum_{\rho=3}^p
           \alpha_{\mu+\rho} + r_{\mu+2} \nonumber \\
  & = & - 2(p-1) - 2 \gamma_{\mu} + 2 \gamma_{\mu+2} + r_{\mu+2},
           \label{eq:r-sol1-1} \\
  r_{\mu+1} & = & 2 + \alpha_{\mu+1} + \alpha_{\mu+2} + r_{\mu+2} = 2 - 2
           \gamma_{\mu+1} + 2 \gamma_{\mu+2} + r_{\mu+2}, \label{eq:r-sol1-2} \\
  r_{\mu+\nu} & = & - 2(\nu-2) - \alpha_{\mu+2} - 2 \sum_{\rho=3}^{\nu-1}
           \alpha_{\mu+\rho} - \alpha_{\mu+\nu} + r_{\mu+2} \nonumber \\
  & = & - 2(\nu-2) + 2 \gamma_{\mu+2} - 2 \gamma_{\mu+\nu} + r_{\mu+2},
           \qquad  \nu = 3, 4, \ldots, p, \label{eq:r-sol1-3}
\end{eqnarray}
where $\gamma_{\mu}$ is defined in Eq.~(\ref{eq:gamma}).\par
%
%
Equation~(\ref{eq:eta-cond1}) restricts the range of $|\eta_{\mu+\nu}|^2$,
$\nu =
1$, 2, \ldots,~$p$, while Eq.~(\ref{eq:eta-cond2}) fixes the value of
$r_{\mu+2}$ in
terms of the latter and the algebra parameters. We conclude that it is
possible to
find values of $\eta_{\mu+\nu}$ and $r_{\nu}$ in Eqs.~(\ref{eq:Q-ansatz})
and~(\ref{eq:H-ansatz}), so that
Eqs.~(\ref{eq:Q-power})--(\ref{eq:multi}) are satisfied. Choosing for instance
\begin{equation}
  |\eta_{\mu+\nu}|^2 = 2, \qquad \nu = 1, 2, \ldots, p,  \label{eq:eta-sol}
\end{equation}
we obtain
\begin{equation}
  r_{\mu+2} = \frac{1}{p} \left[(p-2) \alpha_{\mu+2} + 2 \sum_{\nu=3}^p
(p-\nu+1)
  \alpha_{\mu+\nu} + p (p-2)\right],  \label{eq:r-sol2}
\end{equation}
or
\begin{eqnarray}
  r_{\mu+2} & = & \frac{1}{p} \biggl\{2 \left[1 - (-1)^p\right]
        \sum_{\nu=0}^{\mu+1} (-1)^{\mu+1-\nu} \gamma_{\nu} - 2 \left[p-1 -
        (-1)^p\right] \gamma_{\mu+2} \nonumber\\
  & & \mbox{} + 2 \sum_{\nu=3}^{p-2} \left[1 + (-1)^{p-\nu}\right]
        \gamma_{\mu+\nu} + 4 \gamma_{\mu+p} + p (p-2)\biggr\}.
  \label{eq:r-sol2bis}
\end{eqnarray}
In going from Eq.~(\ref{eq:r-sol2}) to Eq.~(\ref{eq:r-sol2bis}), we used the
inverse of Eq.~(\ref{eq:gamma}), namely
\begin{equation}
  \alpha_{\mu} = \left\{\begin{array}{ll}
        2 \gamma_0, & \mbox{if $\mu=0$}, \\
        4 \sum_{\nu=0}^{\mu-1} (-1)^{\mu-\nu} \gamma_{\nu} + 2\gamma_{\mu}, &
               \mbox{if $\mu=1, 2, \ldots, \lambda-1$}.
  \end{array}\right.
\end{equation}
\par
%
%
{}From Eqs.~(\ref{eq:Hzerobis}), and
(\ref{eq:r-sol1-1})--(\ref{eq:r-sol1-3}), it
follows that the parasupersymmetric Hamiltonian (\ref{eq:H-ansatz}) can be
rewritten as
\begin{equation}
  {\cal H} = N + \frac{1}{2} (2 \gamma_{\mu+2} + r_{\mu+2} - 2p + 3) I +
  \sum_{\nu=1}^p (p + 1 - \nu) P_{\mu+\nu},  \label{eq:H-sol}
\end{equation}
where $r_{\mu+2}$ is given by Eq.~(\ref{eq:r-sol2bis}). The eigenstates
$|n\rangle = |k(p+1) + \nu\rangle$, $n$, $k = 0$, 1, 2,~\ldots, $\nu = 0$, 1,
\ldots,~$p$, of $H_0$ are also eigenstates of $\cal H$, corresponding to the
eigenvalues
\begin{equation}
  {\cal E}_{k(p+1)+\nu} = k (p+1) + \frac{1}{2} (2 \gamma_{\mu+2} + r_{\mu+2} +
  2\mu - 2p + 3), \qquad {\rm if\ } \nu = 0, 1, \ldots, \mu,
  \label{eq:para-spectrum-1}
\end{equation}
\begin{equation}
  {\cal E}_{k(p+1)+\nu} = (k+1) (p+1) + \frac{1}{2} (2 \gamma_{\mu+2} +
r_{\mu+2}
  + 2\mu - 2p + 3), \qquad {\rm if\ } \nu = \mu+1, \mu+2, \ldots, p.
  \label{eq:para-spectrum-2}
\end{equation}
All the levels are therefore equally spaced. The ground state, corresponding to
the energy
\begin{equation}
  {\cal E}_0 = {\cal E}_1 = \cdots = {\cal E}_{\mu} = \frac{1}{2} (2
\gamma_{\mu+2}
  + r_{\mu+2} + 2\mu - 2p + 3),  \label{eq:para-ground}
\end{equation}
is ($\mu+1$)-fold degenerate, whereas the excited states are ($p+1$)-fold
degenerate. Note that since $\mu$ may take any value in the set $\{0, 1, \ldots,
p\}$, the ground-state degeneracy may accordingly vary between 1 and $p+1$.
Unbroken (resp.\ broken) PSSQM corresponds to $\mu=0$ (resp.\ $\mu = 1$, 2,
\ldots, or $p$).\par
%
%
To study the sign of the ground-state energy, we have to insert
Eq.~(\ref{eq:r-sol2bis}) into Eq.~(\ref{eq:para-ground}). The result reads
\begin{eqnarray}
  {\cal E}_0 & = & {\cal E}_1 = \cdots = {\cal E}_{\mu} = \frac{1}{2p} \left[4
  \sum_{\nu=0}^{(\mu-2)/2} \gamma_{2\nu+1} + 4 \sum_{\nu=(\mu+2)/2}^{[p/2]}
  \gamma_{2\nu} + p (2\mu - p + 1)\right], \nonumber\\
  & & {\rm if\ } \mu = 0, 2, \ldots, 2 [p/2], \label{eq:para-groundbis-1}
\end{eqnarray}
\begin{eqnarray}
  {\cal E}_0 & = & {\cal E}_1 = \cdots = {\cal E}_{\mu} = \frac{1}{2p} \left[4
  \sum_{\nu=0}^{(\mu-1)/2} \gamma_{2\nu} + 4 \sum_{\nu=(\mu+1)/2}^{[(p-1)/2]}
  \gamma_{2\nu+1} + p (2\mu - p + 1)\right], \nonumber\\
  & & {\rm if\ } \mu = 1, 3, \ldots, 2 [(p-1)/2] + 1,
  \label{eq:para-groundbis-2}
\end{eqnarray}
where $[a]$ denotes the largest integer contained in $a$, and $\sum_{\nu=a}^b
\equiv 0$ if $a>b$. From the conditions~(\ref{eq:Fock-condbis-1})
and~(\ref{eq:Fock-condbis-2}) for the
existence of
the bosonic Fock-space representation, it follows that
\begin{equation}
  {\cal E}_0 = {\cal E}_1 = \cdots = {\cal E}_{\mu} > \frac{1}{p} (p + 1)
(\mu - p + 1),
  \qquad {\rm if\ } \mu = 0, 1, \ldots, p-2,  \label{eq:ground-bound}
\end{equation}
\begin{equation}
  {\cal E}_0 = {\cal E}_1 = \cdots = {\cal E}_{\mu} > 0, \qquad {\rm if\ }
\mu = p-1, p.
\end{equation}
Since the right-hand side of Eq.~(\ref{eq:ground-bound}) is negative, for $\mu =
0$, 1, \ldots,~$p-2$, the ground-state energy may be positive, null, or negative
according to the values taken by the algebra parameters. We therefore recover a
well-known property of PSSQM of order $p\ge 2$: unlike in SSQM (corresponding to
$p=1$), the energy eigenvalues are not necessarily nonnegative, and there is no
connection between the nonvanishing (resp.\ vanishing) ground-state energy
and the
broken (resp.\ unbroken) PSSQM.\par
%
%
As noted by Khare {\em et al.} (1993b), there is however a special
case in
the standard PSSQM realization
(\ref{eq:standard1})--(\ref{eq:standard-PSSQM2-2}), wherein this
unsatisfactory situation does not
occur,
and moreover the parasupersymmetric Hamiltonian $\cal H$ can be expressed
directly in terms of the parasupercharge operators $Q$, $Q^{\dagger}$, in
contrast
with Eq.~(\ref{eq:multi}). Whenever, in Eq.~(\ref{eq:potential-cond}), all the
constants $C_{\nu}$ vanish, one can indeed write $\cal H$ as
\begin{equation}
  {\cal H} = \frac{1}{2} \left[\left(Q^{\dagger} Q - Q Q^{\dagger}\right)^2 +
  Q^{\dagger} Q^2 Q^{\dagger}\right]^{1/2},  \label{eq:H-special}
\end{equation}
whose eigenvalues are necessarily nonnegative. Furthermore, its ground-state
energy vanishes (resp.\ is positive) for unbroken (resp.\ broken) PSSQM.\par
%
%
Such a special case does have a counterpart in the present bosonic realization.
By introducing Eqs.~(\ref{eq:Q-ansatz}) and~(\ref{eq:H-ansatz}) into
Eq.~(\ref{eq:H-special}), and taking
Eq.~(\ref{eq:eta-sol}) into account, it is easy to show that
Eq.~(\ref{eq:H-special})
is equivalent to the following additional conditions,
\begin{equation}
  r_{\mu} = - 1 - \alpha_{\mu}, \qquad r_{\mu+1} = 1 + \alpha_{\mu+1}, \qquad
  r_{\mu+\nu} = 0, \qquad \nu = 2, 3, \ldots, p,  \label{eq:r-special}
\end{equation}
\begin{equation}
  \alpha_{\mu+\nu} = - 1, \qquad \nu = 2, 3, \ldots, p,
\label{eq:alpha-special}
\end{equation}
which can be checked to be compatible with the previous ones, given in
Eqs.~(\ref{eq:r-sol1-1})--(\ref{eq:r-sol1-3}),
and~(\ref{eq:r-sol2}).\par
%
%
However, the conditions~(\ref{eq:Fock-cond}) for the existence of the bosonic
Fock-space representation are compatible with Eq.~(\ref{eq:alpha-special})
only for
$\mu = 0$ and $\mu = p$. In the former case, $\alpha_1 = p - 1 - \alpha_0$,
$\alpha_2 = \alpha_3 = \cdots = \alpha_p = -1$, where $\alpha_0 > -1$, and from
Eqs.~(\ref{eq:H-sol}) and (\ref{eq:r-special}),
\begin{equation}
  {\cal H} = N + \sum_{\nu=1}^p (p + 1 - \nu) P_{\nu}.
\end{equation}
PSSQM is then unbroken, and the ground-state energy vanishes (${\cal E}_0 =
0$ in
accordance with Eq.~(\ref{eq:para-ground}), since $\gamma_2 = p -
\frac{3}{2}$). In
the latter case, $\alpha_1 = \alpha_2 = \cdots = \alpha_{p-1} = -1$,
$\alpha_p = p -
1 - \alpha_0$, where $\alpha_0 > -1$, and
\begin{equation}
  {\cal H} = N + \sum_{\nu=0}^p (\alpha_0 + 1 - \nu) P_{\nu}.
\end{equation}
PSSQM is then broken, and the ground-state energy ${\cal E}_0 = {\cal E}_1 =
\cdots = {\cal E}_p = \alpha_0 + 1$ (in accordance with
Eq.~(\ref{eq:para-ground}),
since $\gamma_{p+2} = \gamma_1 = \alpha_0 - \frac{1}{2}$) is positive, the
ground
state being ($p+1$)-fold degenerate as all the excited states.\par
%
%
{}Furthermore, by using
conditions~(\ref{eq:r-special}) and~(\ref{eq:alpha-special}), it can be
shown that
$\cal H$ can be rewritten in terms of the supercharges~(\ref{eq:supercharge}) as
\begin{equation}
  {\cal H} = Q_1 Q_1^{\dagger} + \sum_{\nu=1}^p Q_{\nu}^{\dagger} Q_{\nu}.
\end{equation}
This result has also its counterpart in the standard PSSQM
realization (Bagchi {\em et al.}, 1997).\par
%
%
Going back now to the general case corresponding to
conditions~({\ref{eq:r-sol1-1})--(\ref{eq:r-sol2}) only, we note that
Eq.~(\ref{eq:eta-sol}),
yielding the coefficients in the expansion of the
parasupercharges~(\ref{eq:Q-ansatz}), has many solutions. This is not surprising
since Khare did show that in the standard PSSQM
realization~(\ref{eq:standard1})--(\ref{eq:standard-PSSQM2-2}), $\cal H$ has in
fact $p$
(and not only one) conserved parasupercharges, as well as $p$ bosonic
constants (Khare, 1992, 1993). In other words, there exist $p$ independent
operators
$Q_r$, $r=1$, 2, \ldots,~$p$, satisfying with $\cal H$ the set of
equations~(\ref{eq:Q-power})--(\ref{eq:multi}), and $p$ other
independent operators
$I_t$, $t=2$, 3, \ldots,~$p+1$, commuting with $\cal H$, as well as among
themselves. The former are obtained from Eqs.~(\ref{eq:Q-bagchi})
and~(\ref{eq:standard1}) by setting $\sigma_{\nu} = 1$ for $r = 1$, and
$\sigma_{\nu} = 1 - 2 \delta_{\nu,p+1-r}$ for $r = 2$, 3,
\ldots,~$p$, while the latter are given by $(I_t)_{\alpha,\beta} =
\delta_{\alpha,\beta} (1 - 2\delta_{\alpha,t})$, where $t = 2$, 3,
\ldots,~$p+1$,
and $\alpha$, $\beta = 1$, 2, \ldots,~$p+1$. In addition, for any $r_k$,
$r_{k+1}$,
\ldots,~$r_{k+p} \in \{1, 2, \ldots, p\}$,
\begin{equation}
  Q_{r_k} Q_{r_{k+1}} \ldots Q_{r_{k+p}} = 0,  \label{eq:Qr-product}
\end{equation}
and for any $r \in \{1, 2, \ldots, p\}$, $t \in \{2, 3, \ldots, p+1\}$,
\begin{equation}
  [I_t, Q_r] = \sum_{s=1}^p d_{tr}^s Q_s,  \label{eq:It-Qr-com}
\end{equation}
where $d_{tr}^s$ are some real constants, e.g.,
\begin{equation}
  d_{21}^1 = d_{22}^2 = 0, \qquad d_{21}^2 = d_{22}^1 = -2, \qquad d_{31}^1 =
  - d_{31}^2 = - d_{32}^1 = d_{32}^2 = -1, \label{eq:d}
\end{equation}
for $p=2$. Finally, the $Q_r$'s satisfy some mixed multilinear relations
generalizing Eq.~(\ref{eq:multi}), and involving $\cal H$ and the bosonic
constants
$I_t$. For $p=2$, for instance, there are six such independent relations
\begin{equation}
  I_3 Q_s^2 Q_r^{\dagger} + Q_s Q_r^{\dagger} Q_s + I_2 Q_r^{\dagger} Q_s^2 =
  4 Q_r {\cal H}, \label{eq:multi-p=2-1}
\end{equation}
\begin{equation}
  Q_r Q_s Q_s^{\dagger} + Q_s Q_s^{\dagger} Q_r + I_2 Q_s^{\dagger} Q_r Q_s =
  4 Q_r {\cal H}, \label{eq:multi-p=2-2}
\end{equation}
\begin{equation}
  I_3 Q_s Q_r Q_s^{\dagger} + Q_r Q_s^{\dagger} Q_s + Q_s^{\dagger} Q_s Q_r =
  4 Q_r {\cal H}, \label{eq:multi-p=2-3}
\end{equation}
where $(r,s) = (1,2)$, $(2,1)$.\par
%
%
It is straightforward to show that the operators $Q_r$ and $I_t$ have also their
counterpart in the present bosonic realization. Let us indeed consider the
operators
\begin{equation}
  Q_r = \sqrt{2} \sum_{\nu=1}^p b_r^{\nu} a^{\dagger} P_{\mu+\nu}, \qquad r
= 1, 2,
  \ldots, p,  \label{eq:Qr}
\end{equation}
\begin{equation}
  I_t = \sum_{\nu=1}^{p+1} b_t^{\nu} P_{\mu+\nu}, \qquad t = 1, 2, \ldots, p+1,
  \label{eq:It}
\end{equation}
where
\begin{equation}
  b_t^{\nu} = 1 - 2 \delta_{t,\nu} (1 - \delta_{t,1}), \qquad t, \nu = 1,
2, \ldots,
  p+1.
\end{equation}
The $b_t^{\nu}$'s taking values only in the set $\{-1, +1\}$, it is clear
that each
$Q_r$ in Eq.~(\ref{eq:Qr}) satisfies the PSSQM
algebra~(\ref{eq:Q-power})--(\ref{eq:multi}) with
Hamiltonian~(\ref{eq:H-sol}). It is also obvious that $I_2$, $I_3$,
\ldots,~$I_{p+1}$, as defined by Eq.~(\ref{eq:It}), commute with the same,
as well
as among themselves, while $I_1$ reduces to the unit operator.
Equation~(\ref{eq:Qr-product}) directly follows for $n=p$ from the relation
\begin{equation}
  Q_{r_k} Q_{r_{k+1}} \ldots Q_{r_{k+n}} = 2^{(n+1)/2}
\left(a^{\dagger}\right)^{n+1}
  \sum_{\nu=1}^{p-n} B_{\nu}(r_k, r_{k+1}, \ldots, r_{k+n}) P_{\mu+\nu},
\end{equation}
\begin{equation}
  B_{\nu}(r_k, r_{k+1}, \ldots, r_{k+n}) \equiv \prod_{l=0}^n
b_{r_{k+l}}^{\nu+n-l},
\end{equation}
which can be proved by induction over $n$.\par
%
%
Considering now Eq.~(\ref{eq:It-Qr-com}), we obtain from Eqs.~(\ref{eq:Qr})
and~(\ref{eq:It})
\begin{equation}
  [I_t, Q_r] = \sqrt{2}\, a^{\dagger} \sum_{\nu=1}^p c_{tr}^{\nu}
P_{\mu+\nu}, \qquad
  c_{tr}^{\nu} \equiv \left(b_t^{\nu+1} - b_t^{\nu}\right) b_r^{\nu}.
\end{equation}
By combining this result with the inverse of Eq.~(\ref{eq:Qr}),
\begin{equation}
 \sqrt{2}\, a^{\dagger} P_{\mu+\nu} = \sum_{r=1}^p b_{\nu}^r Q_r,
\end{equation}
\begin{equation}
  b_{\nu}^r \equiv \frac{1}{2} \{\delta_{\nu,1} [1 + (2-p) \delta_{r,1}] + (1 -
  \delta_{\nu,1}) (\delta_{r,1} - \delta_{\nu,r})\},
\end{equation}
we get Eq.~(\ref{eq:It-Qr-com}) with $d_{tr}^s$ given by
\begin{equation}
  d_{tr}^s = \sum_{\nu=1}^p c_{tr}^{\nu} b_{\nu}^s.
\end{equation}
For the special cases $p=2$ and $p=3$, considered by Khare (1992, 1993), this
general formula yields the correct results (see e.g. Eq.~(\ref{eq:d})).\par
%
%
{}Finally, for the mixed multilinear relations satisfied by the $Q_r$'s and
$I_t$'s,
let us consider a general relation of the type
\begin{eqnarray}
  & & I_{t_1} Q_{r_1} Q_{r_2} \ldots Q_{r_p} Q_s^{\dagger} + I_{t_2} Q_{r_2}
       Q_{r_3} \ldots Q_{r_p} Q_s^{\dagger} Q_{r_1} + \cdots + I_{t_p} Q_{r_p}
       Q_s^{\dagger} Q_{r_1} Q_{r_2} \ldots Q_{r_{p-1}} \nonumber\\
  & & \mbox{} + I_{t_{p+1}} Q_s^{\dagger} Q_{r_1} Q_{r_2} \ldots Q_{r_p} = 2p
       Q_r^{p-1} {\cal H},  \label{eq:multi-gen}
\end{eqnarray}
where $r_1$, $r_2$, \ldots, $r_p \in \{1, 2, \ldots, p\}$, and $t_1$,
$t_2$, \ldots,
$t_{p+1} \in \{1, 2, \ldots, p+1\}$. It is clear that such a relation
cannot be valid
for any choice of the indices in the ranges indicated. To find to which
choices it
applies when definitions~(\ref{eq:Qr}) and~(\ref{eq:It}) are used, let us work
out the
conditions
implied by Eq.~(\ref{eq:multi-gen}).\par
%
%
After some calculations, one gets
\begin{equation}
  \sum_{\nu=1}^p D_k^{\nu} = p B_k\left([r]^{p-1}\right), \qquad k=1, 2,
  \label{eq:multi-cond1}
\end{equation}
\begin{equation}
  \sum_{\nu=2}^p D_k^{\nu} \left(\nu - 1 + \sum_{\rho=0}^{\nu-2}
  \alpha_{\mu+\rho+2}\right) = \frac{p}{2} B_k\left([r]^{p-1}\right) (1 +
  \alpha_{\mu+2} + r_{\mu+2}), \qquad k=1, 2,  \label{eq:multi-cond2}
\end{equation}
where $[r]^{p-1}$ means that $r$ is repeated ($p-1$) times, and
\begin{equation}
  D_k^{\nu} \equiv b_{t_{\nu+2-k}}^{p+k-1} B_{\nu}(r_{\nu+2-k},
r_{\nu+3-k}, \ldots,
  r_p) b_s^{\nu} B_k(r_1, r_2, \ldots, r_{\nu+1-k}).  \label{eq:D-def}
\end{equation}
Since $B_k\left([r]^{p-1}\right)$ and $D_k^{\nu}$ take values in the set $\{+1,
-1\}$, Eq.~(\ref{eq:multi-cond1}) is satisfied if and only if
\begin{equation}
  D_k^{\nu} = B_k\left([r]^{p-1}\right), \qquad k = 1, 2, \quad \nu = 1, 2,
\ldots, p.
  \label{eq:D-B}
\end{equation}
Then Eq.~(\ref{eq:multi-cond2}) reduces to Eq.~(\ref{eq:eta-cond2}), where the
choice~(\ref{eq:eta-sol}) has been made; hence it is automatically
fulfilled. We are
therefore left with condition~(\ref{eq:D-B}), where we note that
\begin{equation}
  B_1\left([r]^{p-1}\right) = 2 (\delta_{r,1} + \delta_{r,p}) - 1, \qquad
  B_2\left([r]^{p-1}\right) = 2 \delta_{r,1} - 1.  \label{eq:B}
\end{equation}
We conclude that finding all mixed multilinear relations of
type~(\ref{eq:multi-gen}) amounts to determining all sets of $b_t^{\nu}$
coefficients satisfying Eqs.~(\ref{eq:D-def})--(\ref{eq:B}).\par
%
%
Once this has been done, it still remains to eliminate some dependent
relations by
taking into account identities such as
\begin{equation}
  I_{r+1} Q_r = Q_1, \qquad r = 2, 3, \ldots, p,
\end{equation}
\begin{equation}
  I_t Q_1 = Q_{t-1}, \qquad t = 3, 4, \ldots, p+1,
\end{equation}
\begin{equation}
  Q_{r_1} Q_{r_2} \ldots Q_{r_p} Q_s^{\dagger} = I_t Q_{r_1} Q_{r_2} \ldots
Q_{r_p}
  Q_s^{\dagger}, \qquad t = 1, 2, \ldots, p,
\end{equation}
\begin{eqnarray}
  & & Q_{r_k} Q_{r_{k+1}} \ldots Q_{r_p} Q_s^{\dagger} Q_{r_1} Q_{r_2} \ldots
        Q_{r_{k-1}} = I_t Q_{r_k} Q_{r_{k+1}} \ldots Q_{r_p} Q_s^{\dagger}
Q_{r_1}
        Q_{r_2} \ldots Q_{r_{k-1}}, \nonumber\\
  & & \quad k = 2, 3, \ldots, p, \qquad t = 1, 2, \ldots, p-1,
\end{eqnarray}
\begin{equation}
  Q_s^{\dagger} Q_{r_1} Q_{r_2} \ldots Q_{r_p} = I_t Q_s^{\dagger} Q_{r_1}
Q_{r_2}
  \ldots Q_{r_p}, \qquad t = 1, 2, \ldots, p-1, p+1.
\end{equation}
\par
%
%
By proceeding in this way for $p=2$, one gets the six relations given in
Eqs.~(\ref{eq:multi-p=2-1})--(\ref{eq:multi-p=2-3}).
The $p=3$ case can be dealt with in a similar
way, giving
back the results of Khare (1993).\par
%
%
As a final point, let us note that there exists an alternative approach to
PSSQM of
order $p$, due to Beckers and Debergh (1990), wherein
Eq.~(\ref{eq:multi})
is replaced by the cubic equation
\begin{equation}
  \left[Q, \left[Q^{\dagger}, Q\right] \right] = 2Q {\cal H},  \label{eq:cubic}
\end{equation}
while Eqs.~(\ref{eq:Q-power}) and (\ref{eq:H-Q}) remain the same. We proved
elsewhere (Quesne and Vansteenkiste, 1998) that in the $p=2$ case,
Beckers-Debergh PSSQM algebra can only be realized by those ${\cal
A}^{(3)}(G(N))$
algebras that simultaneously bosonize Rubakov-Spiridonov-Khare PSSQM algebra.
For such a reason, we do not consider here that alternative approach to PSSQM of
order
$p$.\par
%
%
\section{\boldmath APPLICATION OF $C_3$-EXTENDED OSCILLATOR ALGEBRAS TO
PSEUDOSSQM}
\label{sec:pseudo}
\setcounter{equation}{0}
PseudoSSQM was introduced by Beckers {\em et al.} (1995a, b) (see also Beckers
and Debergh (1995a, b)) in a study of relativistic vector mesons
interacting with an
external constant magnetic field, wherein the reality of energy eigenvalues was
required. In the nonrelativistic limit,
their theory leads to a pseudosupersymmetric oscillator Hamiltonian, which
can be
realized in terms of bosons and pseudofermions, where the latter are
intermediate
between standard fermions and parafermions of order two. It is then possible to
formulate a pseudoSSQM, characterized by a pseudosupersymmetric Hamiltonian
$\cal H$ and pseudosupercharge operators $Q$, $Q^{\dagger}$, satisfying the
relations
\begin{equation}
  Q^2 = 0,  \label{eq:Q-square}
\end{equation}
\begin{equation}
  [{\cal H}, Q] = 0,  \label{eq:H-Q-pseudo}
\end{equation}
\begin{equation}
  Q Q^{\dagger} Q = 4 c^2 Q {\cal H},  \label{eq:multi-pseudo}
\end{equation}
and their Hermitian conjugates, where $c$ is some real constant. The first two
relations in Eqs.~(\ref{eq:Q-square}), (\ref{eq:H-Q-pseudo}) are the same
as those
occurring in SSQM, whereas the third one in Eq.~(\ref{eq:multi-pseudo}) is
similar
to the multilinear relation valid in PSSQM of order two. Actually, for
$c=1$ or 1/2,
it is compatible with Eq.~(\ref{eq:multi}) or (\ref{eq:cubic}),
respectively.\par
%
%
We will now show that the pseudoSSQM
algebra~(\ref{eq:Q-square})--(\ref{eq:multi-pseudo})
can be
realized in terms of the generators of ${\cal A}^{(3)}(G(N))$ in their bosonic
Fock-space representation. For such a purpose, as in the $p=2$ PSSQM
case (Quesne and Vansteenkiste, 1998), we shall start by assuming
\begin{equation}
  Q = \sum_{\nu=0}^2 \left(\xi_{\nu} a + \eta_{\nu} a^{\dagger}\right) P_{\nu},
  \label{eq:Q-ansatz-pseudo}
\end{equation}
\begin{equation}
  {\cal H} = H_0 + \frac{1}{2} \sum_{\nu=0}^2 r_{\nu} P_{\nu},
  \label{eq:H-ansatz-pseudo}
\end{equation}
where $H_0$ is the bosonic oscillator Hamiltonian~(\ref{eq:Hzero}) associated
with ${\cal A}^{(3)}(G(N))$, $\xi_{\nu}$, $\eta_{\nu}$ are some complex
constants,
and
$r_{\nu}$ some real ones, to be selected in such a way that
Eqs.~(\ref{eq:Q-square})--(\ref{eq:multi-pseudo}) are
satisfied.\par
%
%
Inserting the expression of $Q$, given in Eq.~(\ref{eq:Q-ansatz-pseudo}),
into the
first condition~(\ref{eq:Q-square}), we obtain some restrictions on the
parameters
$\xi_{\nu}$, $\eta_{\nu}$, leading to two sets of three independent
solutions for
$Q$. The solutions belonging to the first set are given by
\begin{equation}
  Q = \left(\xi_{\mu+2} a + \eta_{\mu+2} a^{\dagger}\right) P_{\mu+2},
  \label{eq:Q-interm}
\end{equation}
where $\mu$ takes some fixed, arbitrary value in the set $\{0, 1, 2\}$. Those
belonging to the second set can be written as
\begin{equation}
  Q' = \xi_{\mu+2} a P_{\mu+2} + \eta_{\mu} a^{\dagger} P_{\mu}, \label{eq:Q'}
\end{equation}
and can be obtained from the former by interchanging the roles of $Q$ and
$Q^{\dagger}$ (and changing the $\mu$ value). They will be omitted here, since
$Q$ and $Q^{\dagger}$ play a symmetrical role in the pseudoSSQM
algebra~(\ref{eq:Q-square})--(\ref{eq:multi-pseudo}).\par
%
%
Considering next the second and third conditions in
Eqs.~(\ref{eq:H-Q-pseudo}) and
(\ref{eq:multi-pseudo}), with Q given by Eq.~(\ref{eq:Q-interm}) for some $\mu$
value, and the corresponding $\cal H$ given by
Eq.~(\ref{eq:H-ansatz-pseudo}), we
get the restrictions
\begin{equation}
  \xi_{\mu+2} (- 2 + \alpha_{\mu} + r_{\mu+1} - r_{\mu+2}) = 0,
  \label{eq:pseudo-cond1-1}
\end{equation}
\begin{equation}
  \eta_{\mu+2} (2 - \alpha_{\mu+1} + r_{\mu} - r_{\mu+2}) = 0,
  \label{eq:pseudo-cond1-2}
\end{equation}
and
\begin{equation}
  \left(|\xi_{\mu+2}|^2 + |\eta_{\mu+2}|^2\right) \xi_{\mu+2} = 4 c^2
\xi_{\mu+2}, \label{eq:pseudo-cond2-1}
\end{equation}
\begin{equation}
  \left(|\xi_{\mu+2}|^2 + |\eta_{\mu+2}|^2\right) \eta_{\mu+2} = 4 c^2
\eta_{\mu+2}, \label{eq:pseudo-cond2-2}
\end{equation}
\begin{eqnarray}
  & &\xi_{\mu+2} \left[\left(|\xi_{\mu+2}|^2 + |\eta_{\mu+2}|^2\right) (1 +
       \alpha_{\mu+1}) + |\eta_{\mu+2}|^2 (1 + \alpha_{\mu+2})\right]
\nonumber\\
  & & \mbox{} = 2 c^2 \xi_{\mu+2} (3 + 2 \alpha_{\mu+1} + \alpha_{\mu+2} +
       r_{\mu+2}), \label{eq:pseudo-cond2-3}
\end{eqnarray}
\begin{equation}
  \eta_{\mu+2} |\eta_{\mu+2}|^2 (1 + \alpha_{\mu+2}) = 2 c^2 \eta_{\mu+2} (1 +
  \alpha_{\mu+2} + r_{\mu+2}), \label{eq:pseudo-cond2-4}
\end{equation}
respectively.\par
%
%
Equations~(\ref{eq:pseudo-cond1-1}) and~(\ref{eq:pseudo-cond1-2}) have three
independent solutions:
\begin{equation}
  \xi_{\mu+2} \ne 0, \qquad \eta_{\mu+2} \ne 0, \qquad r_{\mu+1} = 2 -
  \alpha_{\mu} + r_{\mu+2}, \qquad r_{\mu} = - 2 + \alpha_{\mu+1} + r_{\mu+2},
  \label{eq:cond1-sol1}
\end{equation}
\begin{equation}
  \xi_{\mu+2} \ne 0, \qquad \eta_{\mu+2} = 0, \qquad r_{\mu+1} = 2 -
  \alpha_{\mu} + r_{\mu+2}, \label{eq:cond1-sol2}
\end{equation}
\begin{equation}
  \xi_{\mu+2} = 0, \qquad \eta_{\mu+2} \ne 0, \qquad r_{\mu} = - 2 +
  \alpha_{\mu+1} + r_{\mu+2}.
\end{equation}
Since the third solution can be obtained from the second one by substituting
$Q^{\dagger}$ for $Q$, and changing the $\mu$ value, we are only left with
the first
two solutions~(\ref{eq:cond1-sol1}) and~(\ref{eq:cond1-sol2}).\par
%
%
Introducing Eq.~(\ref{eq:cond1-sol1}) into
Eqs.~(\ref{eq:pseudo-cond2-1})--(\ref{eq:pseudo-cond2-4}), we
get the
additional conditions
\begin{equation}
  |\xi_{\mu+2}| = \sqrt{4 c^2 - |\eta_{\mu+2}|^2}, \qquad r_{\mu+2} =
\frac{1}{2c^2}
  (1 + \alpha_{\mu+2}) \left(|\eta_{\mu+2}|^2 - 2 c^2\right),
\label{eq:cond2-sol1}
\end{equation}
which define with Eq.~(\ref{eq:cond1-sol1}) the first set of solutions of the
pseudoSSQM
algebra~(\ref{eq:Q-square})--(\ref{eq:multi-pseudo}). As
we can fix the overall, arbitrary
phase of $Q$ in such a way that $\eta_{\mu+2}$ is real and positive, we
obtain for
each $\mu$ value a two-parameter family of operators
\begin{equation}
  Q(\eta_{\mu+2}, \varphi) = \left(\eta_{\mu+2} a^{\dagger} + e^{{\rm i}
\varphi}
  \sqrt{4 c^2 - \eta_{\mu+2}^2}\, a\right) P_{\mu+2},
\end{equation}
\begin{equation}
  {\cal H}(\eta_{\mu+2}) = N + \frac{1}{2} (2 \gamma_{\mu+2} + r_{\mu+2} - 1) I
  + 2 P_{\mu+1} + P_{\mu+2},
\end{equation}
where $0 < \eta_{\mu+2} < 2 |c|$, $0 \le \varphi < 2\pi$, and $r_{\mu+2}$
is given
by Eq.~(\ref{eq:cond2-sol1}). If we choose for instance $\eta_{\mu+2} = \sqrt{2}
|c|$, and $\varphi = 0$, we get $r_{\mu+2} = 0$, and
\begin{equation}
  Q = c \sqrt{2} \left(a^{\dagger} + a\right) P_{\mu+2},
  \label{eq:pseudo-sol1-1}
\end{equation}
\begin{equation}
  {\cal H} = N + \frac{1}{2} (2 \gamma_{\mu+2} - 1) I + 2 P_{\mu+1} + P_{\mu+2}.
  \label{eq:pseudo-sol1-2}
\end{equation}
Note that this choice does not change $\cal H$ in any significant way since
it only
produces an overall shift of its spectrum.\par
%
%
Introducing now Eq.~(\ref{eq:cond1-sol2}) into
Eqs.~(\ref{eq:pseudo-cond2-1})--(\ref{eq:pseudo-cond2-4})
we get
instead the additional conditions
\begin{equation}
  |\xi_{\mu+2}| = 2 |c|, \qquad r_{\mu+2} = - 1 - \alpha_{\mu+2},
\end{equation}
which define with Eq.~(\ref{eq:cond1-sol2}) a second set of solutions of the
pseudoSSQM
algebra~(\ref{eq:Q-square})--(\ref{eq:multi-pseudo}).
Choosing this time the overall,
arbitrary
phase of $Q$ in such a way that $\xi_{\mu+2}$ is real and positive, we
obtain for
each $\mu$ value a one-parameter family of operators
\begin{equation}
  Q = 2 |c| a P_{\mu+2}, \label{eq:pseudo-sol2-1}
\end{equation}
\begin{equation}
  {\cal H}(r_{\mu}) = N + \frac{1}{2} (2 \gamma_{\mu+2} - \alpha_{\mu+2}) I
  + \frac{1}{2} (1 - \alpha_{\mu+1} + \alpha_{\mu+2} + r_{\mu}) P_{\mu}
  + P_{\mu+1}, \label{eq:pseudo-sol2-2}
\end{equation}
where the parameter $r_{\mu}$ does change the Hamiltonian spectrum in a
significant way.\par
%
%
The pseudosupersymmetric Hamiltonian, corresponding to the first
solution (\ref{eq:pseudo-sol1-1}), (\ref{eq:pseudo-sol1-2}), coincides with
the $p=2$ parasupersymmetric
Hamiltonian previously obtained (Quesne and Vansteenkiste, 1998), and defined
for arbitrary
$p$ in
Eq.~(\ref{eq:H-sol}) of the present work (but the respective charges are of
course
different). Its spectrum and its ground-state energy are therefore given
by Eqs.~(\ref{eq:para-spectrum-1}),~(\ref{eq:para-spectrum-2}),
and by Eq.~(\ref{eq:para-ground}), respectively.\par
%
%
On the contrary, the pseudosupersymmetric Hamiltonian ${\cal H}(r_{\mu})$,
corresponding to the second
solution~(\ref{eq:pseudo-sol2-1}),~(\ref{eq:pseudo-sol2-2}), is new, and its
spectrum is given by
\begin{equation}
  {\cal E}_{3k+\nu} = 3k + \frac{1}{2} (2 \gamma_{\mu+2} - \alpha_{\mu+2} +
  2\mu - 2), \qquad {\rm if\ } \nu = 0, 1, \ldots, \mu-1,
\end{equation}
\begin{equation}
  {\cal E}_{3k+\mu} = 3k + \frac{1}{2} (2 \gamma_{\mu} + r_{\mu} + 2\mu +1),
\end{equation}
\begin{equation}
  {\cal E}_{3k+\nu} = 3k + \frac{1}{2} (2 \gamma_{\mu+2} - \alpha_{\mu+2} +
  2\mu + 4), \qquad {\rm if\ } \nu = \mu+1, \mu+2, \ldots, 2.
\end{equation}
Its levels are therefore equally spaced only if $r_{\mu} = (\alpha_{\mu+1} -
\alpha_{\mu+2} + 3)\, {\rm mod}\, 6$. If $r_{\mu}$ is small enough, the
ground state
is nondegenerate, and its energy is negative for $\mu=1$, or may have any
sign for
$\mu = 0$ or~2. On the contrary, if $r_{\mu}$ is large enough, the ground state
remains nondegenerate with a vanishing energy in the former case, while it
becomes twofold degenerate with a positive energy in the latter. For some
intermediate $r_{\mu}$ value, one gets a two or threefold degenerate ground
state with a vanishing or positive energy, respectively.\par
%
%
\section{\boldmath APPLICATION OF $C_3$-EXTENDED OSCILLATOR ALGEBRAS TO
OSSQM OF ORDER TWO}
\label{sec:ortho}
\setcounter{equation}{0}
OSSQM of arbitrary order $p$ was developed by Khare {\em et
al.} (1993a), by
combining standard bosons with orthofermions of order $p$. The latter had been
previously introduced by Mishra and Rajasekaran (1991a, b), by replacing
Pauli's
exclusion principle by a new, more stringent one. OSSQM is formulated in
terms of
an orthosupersymmetric Hamiltonian $\cal H$, and $p$ orthosupercharge operators
$Q_r$, $Q_r^{\dagger}$, $r = 1$, 2, \ldots,~$p$, satisfying the relations
\begin{equation}
  Q_r Q_s = 0,  \label{eq:ortho-alg1}
\end{equation}
\begin{equation}
  [{\cal H}, Q_r] = 0,  \label{eq:ortho-alg2}
\end{equation}
\begin{equation}
  Q_r Q_s^{\dagger} + \delta_{r,s} \sum_{t=1}^p Q_t^{\dagger} Q_t = 2
\delta_{r,s}
  {\cal H},  \label{eq:ortho-alg3}
\end{equation}
and their Hermitian conjugates, where $r$ and $s$ run over 1, 2,
\ldots,~$p$.\par
%
%
We plan to show that for $p=2$, the OSSQM
algebra~(\ref{eq:ortho-alg1})--(\ref{eq:ortho-alg3})
can be
realized in terms of the generators of ${\cal A}^{(3)}(G(N))$ in their bosonic
Fock-space representation. For such a purpose, let us set
\begin{equation}
  Q_1 = \sum_{\nu=0}^2 \left(\xi_{\nu} a + \eta_{\nu} a^{\dagger}\right)
P_{\nu}, \label{eq:ansatz-ortho-1}
\end{equation}
\begin{equation}
  Q_2 = \sum_{\nu=0}^2 \left(\zeta_{\nu} a + \rho_{\nu} a^{\dagger}\right)
P_{\nu}, \label{eq:ansatz-ortho-2}
\end{equation}
\begin{equation}
  {\cal H} = H_0 + \frac{1}{2} \sum_{\nu=0}^2 r_{\nu} P_{\nu},
  \label{eq:ansatz-ortho-3}
\end{equation}
where we now have at our disposal four types of complex constants $\xi_{\nu}$,
$\eta_{\nu}$, $\zeta_{\nu}$, $\rho_{\nu}$, and one of real ones $r_{\nu}$,
to adjust
in order that Eqs.~(\ref{eq:ortho-alg1})--(\ref{eq:ortho-alg3}) be
satisfied.\par
%
%
Let us first consider Eq.~(\ref{eq:ortho-alg1}) for $r=s=1$, 2. From the study
carried out in Section~\ref{sec:pseudo}, we know that for each $r$ in the set
$\{1, 2\}$,
the equation $Q_r^2 = 0$ admits two different types of solutions, given in
Eqs.~(\ref{eq:Q-interm}) and (\ref{eq:Q'}), respectively, and connected by the
symmetry $Q \leftrightarrow Q^{\dagger}$. In the present case, we have to
distinguish them, since the OSSQM
algebra~(\ref{eq:ortho-alg1})--(\ref{eq:ortho-alg3})
is not invariant
under such a symmetry. Hence, for the couple of orthosupersymmetric charges
$(Q_1, Q_2)$, we get seven types of solutions of $Q_1^2 = Q_2^2 = 0$,
namely $Q_1$
and $Q_2$ may be both of type $Q$, or $Q'$, with the same or adjacent $\mu$
values, or $Q_1$ is of type $Q$ corresponding to a given $\mu$ value, and
$Q_2$ of
type $Q'$ corresponding to $\mu$, $\mu+1$, or $\mu+2$. Here, we take into
account
the fact that the
algebra~(\ref{eq:ortho-alg1})--(\ref{eq:ortho-alg3})
is invariant under the exchange
$Q_1 \leftrightarrow Q_2$.\par
%
%
Imposing next Eqs.~(\ref{eq:ortho-alg1}) and~(\ref{eq:ortho-alg3}) for $r
\ne s$,
i.e., $Q_1 Q_2 = Q_2 Q_1 = Q_1 Q_2^{\dagger} = 0$, we obtain that those
seven cases
for $(Q_1, Q_2)$ actually reduce to two, given by
\begin{equation}
  Q_1 = \xi_{\mu+2} a P_{\mu+2} + \eta_{\mu} a^{\dagger} P_{\mu}, \qquad
  Q_2 = \zeta_{\mu+2} a P_{\mu+2} + \rho_{\mu} a^{\dagger} P_{\mu},
  \label{eq:ortho-sol1}
\end{equation}
and
\begin{equation}
  Q_1 = \xi_{\mu+2} a P_{\mu+2}, \qquad Q_2 = \rho_{\mu} a^{\dagger} P_{\mu},
  \label{eq:ortho-sol2}
\end{equation}
respectively, where for the first one, we have the additional conditions
\begin{equation}
  \xi_{\mu+2} \zeta_{\mu+2}^* + \eta_{\mu} \rho_{\mu}^* = 0 \qquad (\xi_{\mu+2},
  \eta_{\mu} \ne 0),  \label{eq:ortho-cond1}
\end{equation}
\begin{equation}
  \alpha_{\mu+1} = - 1.  \label{eq:ortho-cond2}
\end{equation}
Note that the latter is compatible with conditions~(\ref{eq:Fock-cond}) for the
existence of the bosonic Fock-space representation only for $\mu=0$ and
$\mu=1$.\par
%
%
Equation~(\ref{eq:ortho-alg2}) now leads to the same conditions for both
choices~(\ref{eq:ortho-sol1}) and~(\ref{eq:ortho-sol2}), namely
\begin{equation}
  r_{\mu} = 4 + \alpha_{\mu+1} + r_{\mu+2}, \qquad r_{\mu+1} = 2 - \alpha_{\mu}
  + r_{\mu+2}.  \label{eq:ortho-cond3}
\end{equation}
\par
%
%
It only remains to impose Eq.~(\ref{eq:ortho-alg3}) for $r=s=1$, 2. For the
first
couple of operators $(Q_1, Q_2)$, given in
Eqs.~(\ref{eq:ortho-sol1}),~(\ref{eq:ortho-cond1}),
and~(\ref{eq:ortho-cond2}), we obtain the additional restrictions
\begin{equation}
  |\xi_{\mu+2}|^2 + |\eta_{\mu}|^2 = 2, \qquad |\zeta_{\mu+2}|^2 =
|\eta_{\mu}|^2,
  \qquad |\rho_{\mu}|^2 = |\xi_{\mu+2}|^2, \qquad \xi_{\mu+2} \eta_{\mu}^* +
  \zeta_{\mu+2} \rho_{\mu}^* = 0,  \label{eq:ortho-cond4}
\end{equation}
and
\begin{equation}
  r_{\mu} = 1 + \alpha_{\mu}, \qquad r_{\mu+1} = 0, \qquad r_{\mu+2} = - 1 -
  \alpha_{\mu+2}.   \label{eq:ortho-cond5}
\end{equation}
Combining Eqs.~(\ref{eq:ortho-cond1}) and~(\ref{eq:ortho-cond4}), we get
\begin{equation}
  \xi_{\mu+2} = |\xi_{\mu+2}| e^{{\rm i} \alpha}, \qquad \eta_{\mu} = \sqrt{2 -
  |\xi_{\mu+2}|^2}\, e^{{\rm i} \beta},
\end{equation}
\begin{equation}
  \zeta_{\mu+2} = - \sqrt{2 -
  |\xi_{\mu+2}|^2}\, e^{{\rm i} (\alpha - \beta + \gamma)}, \qquad \rho_{\mu} =
  |\xi_{\mu+2}| e^{{\rm i} \gamma},
\end{equation}
where $0 < |\xi_{\mu+2}| < \sqrt{2}$, and $0 \le \alpha, \beta, \gamma <
2\pi$. In
addition, we find that Eqs.~(\ref{eq:ortho-cond2}), (\ref{eq:ortho-cond3}), and
(\ref{eq:ortho-cond5}) are compatible, and can be combined into the relations
\begin{equation}
  r_{\mu} = 1 + \alpha_{\mu}, \qquad r_{\mu+1} = 0, \qquad r_{\mu+2} = - 2 +
  \alpha_{\mu}, \qquad \alpha_{\mu+1} = - 1.
\end{equation}
Choosing the overall, arbitrary phases of $Q_1$ and $Q_2$ in such a way that
$\xi_{\mu+2}$ and $\rho_{\mu}$ are real and positive, and setting $\beta =
\varphi$, we obtain, for $\mu=0$ or 1, a two-parameter family of solutions of
Eqs.~(\ref{eq:ortho-alg1})--(\ref{eq:ortho-alg3}),
\begin{equation}
  Q_1(\xi_{\mu+2}, \varphi) = \xi_{\mu+2} a P_{\mu+2} + e^{{\rm i} \varphi}
  \sqrt{2 - \xi_{\mu+2}^2}\, a^{\dagger} P_{\mu}, \label{eq:ortho-sol-1}
\end{equation}
\begin{equation}
  Q_2(\xi_{\mu+2}, \varphi) = - e^{-{\rm i} \varphi} \sqrt{2 - \xi_{\mu+2}^2}\,
  a P_{\mu+2} + \xi_{\mu+2} a^{\dagger} P_{\mu}, \label{eq:ortho-sol-2}
\end{equation}
\begin{equation}
  {\cal H} = N + \frac{1}{2} (2 \gamma_{\mu+1} - 1) I + 2 P_{\mu} + P_{\mu+1},
  \label{eq:ortho-H}
\end{equation}
where $0 < \xi_{\mu+2} < \sqrt{2}$, $0 \le \varphi <2\pi$, and $\alpha_{\mu+1} =
-1$.\par
%
%
{}For the second couple of operators $(Q_1, Q_2)$, given in
Eq.~(\ref{eq:ortho-sol2}), Eq.~(\ref{eq:ortho-alg3}) with $r=s=1$, 2 leads
to the
conditions
\begin{equation}
  |\xi_{\mu+2}|^2 = |\rho_{\mu}|^2 = 2,
\end{equation}
and to Eqs.~(\ref{eq:ortho-cond2}) and~(\ref{eq:ortho-cond5}). Hence, with an
appropriate choice of phases, we obtain
Eqs.~(\ref{eq:ortho-sol-1})--(\ref{eq:ortho-H}) with
$\xi_{\mu+2}
= \sqrt{2}$. We conclude that the most general solution of the OSSQM
algebra~(\ref{eq:ortho-alg1})--(\ref{eq:ortho-alg3}) that
can be written in the
form~(\ref{eq:ansatz-ortho-1})--(\ref{eq:ansatz-ortho-3}) is given by
Eqs.~(\ref{eq:ortho-sol-1})--(\ref{eq:ortho-H}),
where
$\mu \in
\{0, 1\}$, $0 < \xi_{\mu+2} \le \sqrt{2}$, $0 \le \varphi < 2\pi$, and
$\alpha_{\mu+1} = -1$.\par
%
%
The orthosupersymmetric Hamiltonian $\cal H$ in Eq.~(\ref{eq:ortho-H}) is
independent of the parameters $\xi_{\mu+2}$, $\varphi$. All the levels of its
spectrum are equally spaced. For $\mu=0$, they are threefold degenerate, since
\begin{equation}
  {\cal E}_{3k} = {\cal E}_{3k+1} = {\cal E}_{3k+2} = 3k + \frac{1}{2} (2
\gamma_1
  + 3).
\end{equation}
OSSQM is therefore broken, and the ground-state energy
\begin{equation}
  {\cal E}_0 = {\cal E}_1 = {\cal E}_2 = \frac{1}{2} (2 \gamma_1 + 3) =
\alpha_0 +
  1
\end{equation}
is positive. On the contrary, for $\mu=1$, only the excited states are threefold
degenerate, since
\begin{equation}
  {\cal E}_{3(k+1)} = {\cal E}_{3k+1} = {\cal E}_{3k+2} = 3k + \frac{1}{2} (2
  \gamma_2 + 5).
\end{equation}
OSSQM is then unbroken, and the ground-state energy
\begin{equation}
  {\cal E}_0 = \frac{1}{2} (2 \gamma_2 - 1) = - \frac{1}{2} (\alpha_2 + 1)
\end{equation}
vanishes. Such results agree with the general conclusions of Khare {\em et al.}
(1993a).\par
%
%
{}For $p$ values greater than two, the OSSQM
algebra~(\ref{eq:ortho-alg1})--(\ref{eq:ortho-alg3})
becomes
rather complicated because the number of equations to be fulfilled increases
considerably. A glance at the 18 independent conditions for $p=3$ led us to the
conclusion that the ${\cal A}^{(4)}(G(N))$ algebra is not rich enough to contain
operators satisfying Eqs.~(\ref{eq:ortho-alg1})--(\ref{eq:ortho-alg3}).
Contrary to
what happens for PSSQM,
for OSSQM the $p=2$ case is therefore not representative of the general one.\par
%
%
\section{\boldmath SOME DEFORMED $C_{\lambda}$-EXTENDED OSCILLATOR
ALGEBRAS}
\label{sec:deformations}
\setcounter{equation}{0}
The purpose of the present section is to construct some deformations of the
$C_{\lambda}$-extended oscillator algebras ${\cal A}^{(\lambda)}_{\alpha_0
\alpha_1 \ldots \alpha_{\lambda-2}}$, subject to the condition that they admit
three Casimir operators analogous to ${\cal C}_1$, ${\cal C}_2$, ${\cal C}_3$,
defined in Eqs.~(\ref{eq:casimir1})--(\ref{eq:casimir3}).\par
%
%
Let us consider a class of algebras generated by $I$, $a^{\dagger}$, $a =
\left(a^{\dagger}\right)^{\dagger}$, $N = N^{\dagger}$,
$P_{\mu}^{\vphantom{\dagger}} = P_{\mu}^{\dagger}$, $\mu=0$, 1,
$\ldots$,~$\lambda-1$, satisfying the defining
relations~(\ref{eq:alg-def2-1})--(\ref{eq:alg-def2-3})
of
${\cal A}^{(\lambda)}_{\alpha_0 \alpha_1 \ldots \alpha_{\lambda-2}}$, except for
the commutator of $a$ and $a^{\dagger}$ in Eq.~(\ref{eq:alg-def2-3}), which is
replaced by the quommutator (or $q$-deformed commutator)
\begin{equation}
  \left[a, a^{\dagger}\right]_q \equiv a a^{\dagger} - q a^{\dagger} a =
H(N) + K(N)
  \sum_{\mu=0}^{\lambda-1} \alpha_{\mu} P_{\mu},  \label{eq:quom}
\end{equation}
where $q \in \R^+$, $\alpha_{\mu} \in \R$, and $H(N)$, $K(N)$
are some
real, analytic functions of $N$.\par
%
%
The operators ${\cal  C}_1$, ${\cal  C}_2$ of Eqs.~(\ref{eq:casimir1}),
(\ref{eq:casimir2}) remain invariants of the new algebras. We will determine the
constraints that the existence of a third Casimir operator of the type
\begin{equation}
  \tilde{\cal C}_3 = q^{-N} \left(D(N) + E(N) \sum_{\mu=0}^{\lambda-1}
\beta_{\mu}
  P_{\mu} - a^{\dagger} a \right)
\end{equation}
imposes on $H(N)$ and $K(N)$, assuming that Eq.~(\ref{eq:alpha-cond}) is
the only
relation satisfied by the $\alpha_{\mu}$'s. Here $\beta_{\mu}$, $\mu=0$, 1,
\ldots,~$\lambda - 1$, and $D(N)$, $E(N)$ are assumed to be some real constants,
and some real, analytic functions of $N$, respectively. In the case of the
undeformed
algebras ${\cal A}^{(\lambda)}_{\alpha_0 \alpha_1 \ldots \alpha_{\lambda-2}}$,
one has $q=1$, $H(N) = K(N) = I$, and $\tilde{\cal C}_3$ reduces to ${\cal
C}_3$,
given in Eq.~(\ref{eq:casimir3}), with $D(N) = N$, $E(N) = I$, and $\beta_{\mu}$
defined by Eq.~(\ref{eq:beta}) in terms of the $\alpha_{\mu}$'s.\par
%
%
In the realization~(\ref{eq:T-GDOA}), the deformed algebras, defined by
Eqs.~(\ref{eq:alg-def2-1}), (\ref{eq:alg-def2-2}), (\ref{eq:alpha-cond}), and
(\ref{eq:quom}), reduce to GDOAs ${\cal A}^{(\lambda)}_q(G(N))$, with $q
\ne 1$ and
$G(N)$ given by the right-hand side of Eq.~(\ref{eq:quom}). Then
$\tilde{\cal C}_3$
reduces to the standard Casimir operator $\tilde{\cal C}$ of such algebras,
and $F(N)
= D(N) + E(N) \sum_{\mu=0}^{\lambda-1} \beta_{\mu} P_{\mu}$ becomes the GDOA
structure function, satisfying the equation $F(N+1) - q F(N) =
G(N)$ (Katriel and Quesne, 1996; Quesne and Vansteenkiste, 1996, 1997).\par
%
%
Going back to the general case, we note that since $\tilde{\cal C}_3$ is a
Hermitian
operator commuting with $N$ and $P_{\mu}$, we only have to impose the condition
$\left[\tilde{\cal C}_3, a\right] = 0$. By using the defining relations, it
is easy to
show that the latter is equivalent to the two functional equations
\begin{equation}
  D(N+1) - q D(N) = H(N),  \label{eq:funct1}
\end{equation}
\begin{equation}
  E(N+1) \beta_{\mu+1} - q E(N) \beta_{\mu} = K(N) \alpha_{\mu}, \qquad \mu
= 0, 1,
  \ldots, \lambda - 1,  \label{eq:funct2}
\end{equation}
where we assume as usual $\beta_{\lambda} = \beta_0$. Equation~(\ref{eq:funct1})
is similar to the equation appearing in the construction of $\tilde{\cal C}$ for
GDOAs with $q\ne 1$ (Katriel and Quesne, 1996; Quesne and Vansteenkiste, 1996,
1997), while Eq.~(\ref{eq:funct2}) is
a new
functional equation, whose solutions will now be determined.\par
%
%
{}For such a purpose, let us consider the following nonhomogeneous system of
$\lambda$ linear equations in $\lambda$ unknowns $\beta_{\mu}$, $\mu = 0$, 1,
\ldots,~$\lambda-1$,
\begin{equation}
  - q E(x) \beta_{\mu} + E(x+1) \beta_{\mu+1} = K(x) \alpha_{\mu}, \qquad
\mu = 0,
1,
  \ldots, \lambda - 1, \label{eq:beta-system-1}
\end{equation}
\begin{equation}
  \beta_{\lambda} \equiv \beta_0, \label{eq:beta-system-2}
\end{equation}
where $x$ is some real variable.\par
%
%
If the determinant of its coefficient matrix is nonvanishing, i.e., if
\begin{equation}
  [E(x+1)]^{\lambda} - [q E(x)]^{\lambda} \ne 0,
\end{equation}
or, equivalently,
\begin{equation}
  E(x) \ne b q^x,  \label{eq:E-cond1}
\end{equation}
and
\begin{equation}
  E(x) \ne b' (-q)^x, \qquad \mbox{\rm if $\lambda$ is even},
\label{eq:E-cond2}
\end{equation}
where $b$, $b'$ are some real, nonvanishing constants, then the system has
one and
only one solution, given by
\begin{equation}
  \beta_{\mu} = \frac{[q E(x)]^{\lambda-1} K(x)}{[E(x+1)]^{\lambda} - [q
  E(x)]^{\lambda}} \sum_{\nu=0}^{\lambda-1}
\left(\frac{E(x+1)}{qE(x)}\right)^{\nu}
  \alpha_{\mu+\nu}, \qquad \mu=0, 1, \ldots, \lambda-1. \label{eq:unique-sol}
\end{equation}
Since, by definition, $\beta_{\mu}$, $\mu=0$, 1, \ldots,~$\lambda-1$, are
constants, the functions $E(x)$ and $K(x)$ should be chosen in such a way
that the
dependence on $x$ disappears on the right-hand side of
Eq.~(\ref{eq:unique-sol}).\par
%
%
Let us first consider $\beta_0$. By using Eq.~(\ref{eq:alpha-cond}) to express
$\alpha_0$ in terms of $\alpha_1$, $\alpha_2$, \ldots,~$\alpha_{\lambda-1}$,
$\beta_0$ can be rewritten as
\begin{equation}
  \beta_0 = \frac{[q E(x)]^{\lambda-1} K(x)}{[E(x+1)]^{\lambda} - [q
  E(x)]^{\lambda}} \sum_{\nu=1}^{\lambda-1} \left[\left(
  \frac{E(x+1)}{qE(x)}\right)^{\nu} - 1\right] \alpha_{\nu}.
\label{eq:beta-zero}
\end{equation}
Since $\alpha_1$, $\alpha_2$, \ldots,~$\alpha_{\lambda-1}$ are assumed to be
independent, the coefficient of each of them on the right-hand side of
Eq.~(\ref{eq:beta-zero}) should reduce to some real constant, which we denote by
$e_{\nu}$, $\nu=1$, 2, \ldots,~$\lambda-1$. Hence we get the system of equations
\begin{equation}
  \frac{1}{e_1} \left(\frac{E(x+1)}{qE(x)} - 1\right) =
\frac{[E(x+1)]^{\lambda} - [q
  E(x)]^{\lambda}}{[q E(x)]^{\lambda-1} K(x)},  \label{eq:E-K1}
\end{equation}
\begin{equation}
  \frac{1}{e_1} \left(\frac{E(x+1)}{qE(x)} - 1\right) = \frac{1}{e_{\nu}}
\left[\left(
  \frac{E(x+1)}{qE(x)}\right)^{\nu} - 1\right], \qquad \nu=2, 3, \ldots,
\lambda-1,
  \label{eq:E-K2}
\end{equation}
to determine the constraints on $E(x)$ and $K(x)$.\par
%
%
{}For $\lambda=2$, we are only left with the first
equation~(\ref{eq:E-K1}), yielding
the constraint
\begin{equation}
  K(x) = e_1 [E(x+1) + q E(x)].
\end{equation}
Introducing the latter into Eq.~(\ref{eq:unique-sol}), and using
Eq.~(\ref{eq:alpha-cond}) again, we obtain
\begin{equation}
  \beta_{\mu} = - e_1 \alpha_{\mu}, \qquad \mu = 0, 1,
\end{equation}
which are constants as it should be. Incorporating the constant $e_1$ into
the $E(x)$
definition, we conclude that the algebras defined by Eqs.~(\ref{eq:alg-def2-1}),
(\ref{eq:alg-def2-2}) with $\lambda=2$, and
\begin{equation}
  \left[a, a^{\dagger}\right]_q = H(N) + [E(N+1) + q E(N)] (\alpha_0 P_0 +
\alpha_1
  P_1),  \label{eq:defalg1-1}
\end{equation}
where $\alpha_0$, $\alpha_1$ satisfy Eq.~(\ref{eq:alpha-cond}), $H(N)$ is
arbitrary, and $E(N) \ne (\pm q)^N$, admit the three Casimir
operators~(\ref{eq:casimir1}), (\ref{eq:casimir2}), and
\begin{equation}
  \tilde{\cal C}_3 = q^{-N} \left[D(N) - E(N) (\alpha_0 P_0 + \alpha_1 P_1)
  - a^{\dagger} a \right],  \label{eq:defalg1-2}
\end{equation}
where $D(N)$ is some solution of Eq.~(\ref{eq:funct1}). By choosing that
solution
for which $D(0) = \alpha_0 E(0)$, $\tilde{\cal C}_3$ vanishes in the bosonic
Fock-space representation.\par
%
%
{}For $\lambda > 2$, Eq.~(\ref{eq:E-K2}) for $\nu=2$ yields the constraint
\begin{equation}
  E(x+1) = \left(\frac{e_2}{e_1} - 1\right) q E(x),
\end{equation}
whose solution is given by
\begin{equation}
  E(x) = b k^x,
\end{equation}
where $b$ is some real constant, and $k \equiv \left(e_1^{-1} e_2 - 1\right) q$.
{}From Eqs.~(\ref{eq:E-cond1}) and~(\ref{eq:E-cond2}), it follows that for any
$\lambda$, $k \ne q$, and in addition for even $\lambda$, $k \ne - q$.
Equation~(\ref{eq:E-K1}) then provides the expression of $K(x)$,
\begin{equation}
  K(x) = B k^x,
\end{equation}
where $B \equiv e_1 b q^{2-\lambda} \left(k^{\lambda} -
q^{\lambda}\right)/(k-q)$,
while for the remaining $\nu$ values, Eq.~(\ref{eq:E-K2}) leads to the
conditions
\begin{equation}
  e_{\nu} = e_1 q^{1-\nu} \frac{k^{\nu} - q^{\nu}}{k-q}, \qquad \nu = 2, 3,
\ldots,
  \lambda-1.
\end{equation}
Hence, from Eq.~(\ref{eq:unique-sol}), $\beta_{\mu}$ is given by
\begin{equation}
  \beta_{\mu} = \frac{B q^{\lambda-1}}{b(k^{\lambda} - q^{\lambda})}
  \sum_{\nu=0}^{\lambda-1} \left(\frac{k}{q}\right)^{\nu} \alpha_{\mu+\nu},
\qquad
  \mu = 0, 1, \ldots, \lambda-1,
\end{equation}
and therefore reduces to some constant as it should be. We conclude that for
$\lambda>2$, the algebras defined by Eqs.~(\ref{eq:alg-def2-1}),
(\ref{eq:alg-def2-2}), and
\begin{equation}
  \left[a, a^{\dagger}\right]_q = H(N) + B k^N \sum_{\mu=0}^{\lambda-1}
  \alpha_{\mu} P_{\mu},  \label{eq:defalg2-1}
\end{equation}
where $H(N)$ and $B$ are arbitrary, $\alpha_{\mu}$ satisfies
Eq.~(\ref{eq:alpha-cond}), $k \ne q$ for any $\lambda$, and $k \ne -q$ for even
$\lambda$, admit the three Casimir operators~(\ref{eq:casimir1}),
(\ref{eq:casimir2}), and
\begin{equation}
  \tilde{\cal C}_3 = q^{-N} \left\{D(N) + \frac{B q^{\lambda-1}}{k^{\lambda} -
  q^{\lambda}} k^N \sum_{\mu=0}^{\lambda-1} \left[\sum_{\nu=0}^{\lambda-1}
  \left(\frac{k}{q}\right)^{\nu} \alpha_{\mu+\nu}\right] P_{\mu}
  - a^{\dagger} a \right\},  \label{eq:defalg2-2}
\end{equation}
where $D(N)$ is some solution of Eq.~(\ref{eq:funct1}). By choosing that
solution
for which $D(0) = - B q^{\lambda-1} \left(k^{\lambda} - q^{\lambda}\right)^{-1}
\sum_{\nu=0}^{\lambda-1} (k/q)^{\nu} \alpha_{\nu}$, $\tilde{\cal C}_3$
vanishes in
the bosonic Fock-space representation.\par
%
%
It remains to consider the cases where the coefficient matrix of
system~(\ref{eq:beta-system-1}),~(\ref{eq:beta-system-2}) has a vanishing
determinant. If $E(x) = b
q^x$, where
$b$ is some real constant, then Eqs.~(\ref{eq:beta-system-1})
and~(\ref{eq:beta-system-2}) become
\begin{equation}
  - \beta_{\mu} + \beta_{\mu+1} = (bq)^{-1} \frac{K(x)}{q^x} \alpha_{\mu},
\qquad
  \mu = 0, 1, \ldots, \lambda - 1,
\end{equation}
\begin{equation}
  \beta_{\lambda} \equiv \beta_0.
\end{equation}
Since the $\beta_{\mu}$'s are constants, we obtain
\begin{equation}
  K(x) = B q^x,
\end{equation}
where $B$ is some real constant, and therefore
\begin{equation}
  \beta_{\mu} = \frac{B}{bq} \sum_{\nu=0}^{\mu-1} \alpha_{\nu} + \beta_0, \qquad
  \mu = 1, 2, \ldots, \lambda - 1.
\end{equation}
We conclude that the algebras defined by Eqs.~(\ref{eq:alg-def2-1}),
(\ref{eq:alg-def2-2}), and
\begin{equation}
  \left[a, a^{\dagger}\right]_q = H(N) + B q^N \sum_{\mu=0}^{\lambda-1}
  \alpha_{\mu} P_{\mu},  \label{eq:defalg3-1}
\end{equation}
where $H(N)$ and $B$ are arbitrary, and $\alpha_{\mu}$ satisfies
Eq.~(\ref{eq:alpha-cond}), admit the three Casimir
operators (\ref{eq:casimir1}),
(\ref{eq:casimir2}), and
\begin{equation}
  \tilde{\cal C}_3 = q^{-N} \left[D(N) + B q^{N-1} \sum_{\mu=1}^{\lambda-1}
  \left(\sum_{\nu=0}^{\mu-1} \alpha_{\nu}\right) P_{\mu}
  - a^{\dagger} a \right],  \label{eq:defalg3-2}
\end{equation}
where we have set $\beta_0 = 0$ (thereby eliminating a multiple of the unit
operator), and $D(N)$ is some solution of Eq.~(\ref{eq:funct1}). By
choosing that
solution for which $D(0) = 0$, $\tilde{\cal C}_3$ vanishes in the bosonic
Fock-space
representation.\par
%
%
{}Finally, if $\lambda$ is even, and $E(x) = b (-q)^x$, where $b$ is some real
constant, then Eqs.~(\ref{eq:beta-system-1}) and~(\ref{eq:beta-system-2}) become
\begin{equation}
  \beta_{\mu} + \beta_{\mu+1} = - (bq)^{-1} \frac{K(x)}{(-q)^x}
\alpha_{\mu}, \qquad
  \mu = 0, 1, \ldots, \lambda - 1,  \label{eq:beta-1}
\end{equation}
\begin{equation}
  \beta_{\lambda} \equiv \beta_0.  \label{eq:beta-2}
\end{equation}
The $\beta_{\mu}$ constancy implies again that
\begin{equation}
  K(x) = B (-q)^x,
\end{equation}
where $B$ is some real constant. Equation~(\ref{eq:beta-1}) is then
equivalent to
\begin{equation}
  \beta_0 + \beta_1 = - \frac{B}{bq} \alpha_0, \label{eq:beta-1bis-1}
\end{equation}
\begin{equation}
  \beta_{\mu+2} - \beta_{\mu} = - \frac{B}{bq} (\alpha_{\mu+1} - \alpha_{\mu}),
  \qquad \mu = 0, 1, \ldots, \lambda - 2. \label{eq:beta-1bis-2}
\end{equation}
The solution of Eqs.~(\ref{eq:beta-1bis-1}) and~(\ref{eq:beta-1bis-2}) is given
by
\begin{equation}
  \beta_{\mu} = - \frac{B}{bq} \left(\sum_{\nu=0}^{(\mu-2)/2} \alpha_{2\nu+1} -
  \sum_{\nu=0}^{(\mu-2)/2} \alpha_{2\nu} \right) + \beta_0, \qquad \mbox{\rm if
  $\mu$ is even},  \label{eq:beta-even}
\end{equation}
\begin{equation}
  \beta_{\mu} = - \frac{B}{bq} \left(\sum_{\nu=0}^{(\mu-1)/2} \alpha_{2\nu} -
  \sum_{\nu=0}^{(\mu-3)/2} \alpha_{2\nu+1} \right) - \beta_0, \qquad
\mbox{\rm if
  $\mu$ is odd}.
\end{equation}
Condition~(\ref{eq:beta-2}) is consistent with Eq.~(\ref{eq:beta-even}) if and
only if we impose that $\sum_{\nu=0}^{(\lambda-2)/2} \alpha_{2\nu+1} =
\sum_{\nu=0}^{(\lambda-2)/2} \alpha_{2\nu}$, or by taking
Eq.~(\ref{eq:alpha-cond}) into account, $\sum_{\nu=0}^{(\lambda-2)/2}
\alpha_{2\nu} = 0$. Since we have assumed that the $\alpha_{\mu}$'s do not
satisfy any extra relation apart from Eq.~(\ref{eq:alpha-cond}), the case
$E(x) =
b (-q)^x$ has to be rejected.\par
%
%
We therefore found altogether three deformed $C_{\lambda}$-extended
oscillator algebras admitting three Casimir operators ${\cal C}_1$, ${\cal
C}_2$, $\tilde{\cal C}_3$. They correspond to Eqs.~(\ref{eq:defalg1-1})
and~(\ref{eq:defalg1-2}), (\ref{eq:defalg2-1}) and~(\ref{eq:defalg2-2}),
(\ref{eq:defalg3-1}) and~(\ref{eq:defalg3-2}), respectively.\par
%
%
The deformed Calogero-Vasiliev algebra introduced by Brzezi\'nski {\em et
al.}~(1993), for which
\begin{equation}
  \left[a, a^{\dagger}\right]_q = q^{-N} (1 + 2\alpha K), \qquad K = (-1)^N,
\end{equation}
is a special case of Eq.~(\ref{eq:defalg1-1}), corresponding to
\begin{equation}
  H(N) = q^{-N}, \qquad E(N) = \frac{2q^{-N}}{q+q^{-1}}, \qquad \alpha_0 =
- \alpha_1
  = \alpha.
\end{equation}
{}From Eq.~(\ref{eq:funct1}), we obtain
\begin{equation}
  D(N) = \frac{q^N-q^{-N}}{q-q^{-1}} + \frac{2\alpha q^N}{q+q^{-1}},
\end{equation}
so that the Casimir operator~(\ref{eq:defalg1-2}) becomes
\begin{equation}
  \tilde{\cal C}_3 = q^{-N} \left(\frac{q^N-q^{-N}}{q-q^{-1}} +
\frac{2\alpha (q^N -
  q^{-N} K)}{q+q^{-1}} - a^{\dagger} a \right).  \label{eq:def-CV}
\end{equation}
In a given unirrep, whose basis states are given by
Eq.~(\ref{eq:unnormalized}), and
satisfy relations similar to Eqs.~(\ref{eq:unnormalized-rel-1})
and~(\ref{eq:unnormalized-rel-2}) with ${\cal C}
= {\cal
C}_3$ replaced by $\tilde{\cal C}_3$, we obtain from Eq.~(\ref{eq:def-CV}) that
$\lambda_n$ can be expressed as
\begin{equation}
  \lambda_n = - q^{n_0+n} c + \frac{q^{n_0+n}-q^{-n_0-n}}{q-q^{-1}} + 2\alpha
  \frac{q^{n_0+n}-(-q)^{-n_0-n}}{q+q^{-1}},
\end{equation}
or
\begin{equation}
  \lambda_n = q^n \lambda_0 + q^{-n_0} \left(\frac{q^n-q^{-n}}{q-q^{-1}} + B
  \frac{q^n-(-q)^{-n}}{q+q^{-1}}\right), \qquad B \equiv 2\alpha (-1)^{n_0}.
\end{equation}
This equation is consistent with Eq.~(14) of Kosi\'nski {\em et al.} (1997),
wherein
the representations of the deformed Calogero-Vasiliev algebra were studied. Note
that this result holds although there are some slight discrepancies in the
algebra
definition between Kosi\'nski {\em et al.} (1997) and the present work, and the
Casimir operator $\tilde{\cal C}_3$ was not considered in the former.\par
%
%
Some interesting special cases of the algebras corresponding to
Eqs.~(\ref{eq:defalg2-1}) and~(\ref{eq:defalg3-1}) are obtained for $q=1$,
$k \ne
1$, $k \ne -1$ (if $\lambda$ is even), and $k=1$, $q \ne 1$, $q \ne -1$ (if
$\lambda$ is even) for the former, and $q=1$ for the latter.\par
%
%
\section{CONCLUSION}
\label{sec:conclusion}
In the present paper, we studied some mathematical properties of
$C_{\lambda}$-extended oscillator algebras ${\cal A}^{(\lambda)}_{\alpha_0
\alpha_1 \ldots \alpha_{\lambda-2}}$. We constructed Casimir operators, and used
them to provide a complete unirrep classification under the assumption that the
number operator spectrum is nondegenerate. We established that only BFB and FD
unirreps occur, and showed that the unirreps of ${\cal A}^{(\lambda)}_{\alpha_0
\alpha_1 \ldots \alpha_{\lambda-2}}$ can be related to those of its GDOA
realization ${\cal A}^{(\lambda)}(G(N))$.\par
%
%
In addition, we looked for some deformations of ${\cal A}^{(\lambda)}_{\alpha_0
\alpha_1 \ldots \alpha_{\lambda-2}}$, subject to the condition that they admit
Casimir operators analogous to those of the undeformed algebras. We found three
new types of algebras, defined in Eqs.~(\ref{eq:defalg1-1})
and~(\ref{eq:defalg1-2}),
(\ref{eq:defalg2-1}) and~(\ref{eq:defalg2-2}), (\ref{eq:defalg3-1})
and~(\ref{eq:defalg3-2}), respectively. The first one includes the Brzezi\'nski
{\em et al.} (1993) deformation of the Calogero-Vasiliev
algebra (Vasiliev, 1991; Polychronakos, 1992; Brink {\em et al.},
1992; Brink and Vasiliev, 1993) as a special case.\par
%
%
{}Furthermore, we established that the bosonic Fock-space realization of ${\cal
A}^{(\lambda)}(G(N))$ yields a convenient bosonization of several SSQM variants:
PSSQM of order $p = \lambda - 1$ for any $\lambda$, as well as pseudoSSQM, and
OSSQM of order two for $\lambda=3$. In the former case, we provided a full
analysis of the problem, including the construction of the $p$ independent
conserved parasupercharges, and $p$ bosonic constants admitted by the
parasupersymmetric Hamiltonian. Such results  generalize those already known
for standard SSQM (Brzezi\'nski {\em et al.},
1993; Plyushchay, 1996a, b). In the OSSQM case, however, it was not
possible to extend the results to $p$ values greater than two in the
$C_{\lambda}$-extended oscillator algebra context.\par
%
%
There remain some interesting open questions for future study. Apart from those
mentioned in Section~\ref{sec:intro}, we would like to mention here two of
them. The
first one is to further study deformations both from theoretical and applied
viewpoints. Generalizing, for instance, the Macfarlane (1994)
deformation of the Calogero-Vasiliev
algebra
would be an interesting topic. The second issue is to construct some GDOA, whose
structure would be rich enough to enable the OSSQM bosonization to be
carried out
for $p>2$.\par
%
%
\section*{ACKNOWLEDGMENT}

One of the authors (CQ) is a Research Director of the National Fund for
Scientific
Research (FNRS), Belgium. The other (NV) is a Scientific Associate of the
Inter-University Institute for Nuclear Sciences (IISN), Belgium.\par
%
%
\newpage
\setlength{\parindent}{0cm}
\section*{FOOTNOTES}

$^1$In both the oscillator and Heisenberg algebras, the
creation and
annihilation operators $a^{\dagger}$, $a$ are considered as generators, but
in the
former the number operator $N$ appears as an additional independent generator,
whereas in the latter it is defined in terms of $a^{\dagger}$, $a$ as $N \equiv
a^{\dagger} a$.\par
$^2$In a recent study (Guichardet, 1998), the
assumption that the spectrum of $N$ is nondegenerate has been lifted for the
Arik-Coon GDOA (Arik and Coon, 1976; Kuryshkin, 1980), but it has been shown
that this
condition
is automatically fulfilled.\par
%
%
\newpage
\section*{REFERENCES}

Arik, M., and Coon, D. D.\ (1976).\ {\em Journal of Mathematical Physics}, {\bf
17}, 524.

Bagchi, B.\ (1994).\ {\em Physics Letters A}, {\bf 189}, 439.

Bagchi, B., Biswas, S. N., Khare, A., and Roy, P. K.\ (1997).\ {\em Pramana
- Journal
of Physics}, {\bf 49}, 199.

Beckers, J., and Debergh, N.\ (1990).\ {\em Nuclear Physics B}, {\bf 340}, 767.

Beckers, J., and Debergh, N.\ (1995a).\ {\em International Journal of Modern
Physics A}, {\bf 10}, 2783.

Beckers, J., and Debergh, N.\ (1995b).\ In {\em Second International
Workshop on
Harmonic Oscillators, Cocoyoc, Morelos, Mexico, March 23--25, 1994},  D.\
Han and
K.\ B.\ Wolf, eds., NASA Conference Publication 3286, NASA Goddard Space  Flight
Center, Greenbelt, Maryland, p.\ 313.

Beckers, J., Debergh, N., and Nikitin, A. G.\ (1995a).\ {\em Fortschritte der
Physik}, {\bf 43}, 67.

Beckers, J., Debergh, N., and Nikitin, A. G.\ (1995b).\ {\em Fortschritte der
Physik}, {\bf 43}, 81.

Beckers, J., Debergh, N., and Nikitin, A. G.\ (1997).\ {\em International
Journal of Theorical Physics}, {\bf 36}, 1991.

Biedenharn, L. C.\ (1989).\ {\em Journal of Physics A}, {\bf 22}, L873.

Bogoliubov, N. M., Rybin, A. V., and Timonen, J.\ (1994).\
{\em Journal of Physics A}, {\bf 27}, L363.

Bonatsos, D.\ (1992).\ {\em Journal of Physics A}, {\bf 25}, L101.

Bonatsos, D., and Daskaloyannis, C.\ (1992a).\ {\em Physics Letters B}, {\bf
278}, 1.

Bonatsos, D., and Daskaloyannis, C.\ (1992b).\ {\em Physical Review A}, {\bf
46}, 75.

Bonatsos, D., and Daskaloyannis, C.\ (1993a).\ {\em Physics Letters B}, {\bf
307}, 100.

Bonatsos, D., and Daskaloyannis, C.\ (1993b).\ {\em Chemical Physics Letters},
{\bf 203}, 150.

Bonatsos, D., and Daskaloyannis, C.\ (1993c).\ {\em Physical Review A}, {\bf
48}, 3611.

Bonatsos, D., Daskaloyannis, C., and Kokkotas, K.\ (1993).\ {\em Physical
Review A}, {\bf 48}, R3407.

Bonatsos, D., Daskaloyannis, C., and Kokkotas, K.\ (1994).\ {\em Physical
Review A}, {\bf 50}, 3700.

Brink, L., Hansson, T. H., and Vasiliev, M. A.\ (1992).\ {\em Physics
Letters B}, {\bf
286}, 109.

Brink, L., and Vasiliev, M. A.\ (1993).\ {\em Modern Physics Letters A}, {\bf
8}, 3585.

Brzezi\'nski, T., Egusquiza, I. L., and Macfarlane, A. J.\ (1993).\ {\em Physics
Letters B}, {\bf 311}, 202.

Calogero, F.\ (1969a).\ {\em Journal of Mathematical Physics}, {\bf 10}, 2191.

Calogero, F.\ (1969b).\ {\em Journal of Mathematical Physics}, {\bf 10}, 2197.

Calogero, F.\ (1971).\ {\em Journal of Mathematical Physics}, {\bf 12}, 419.

Calogero, F., and Marchioro, C.\ (1974).\ {\em Journal of Mathematical Physics},
{\bf 15}, 1425.

Chang, Z., Guo, H. Y., and Yan, H.\ (1991).\ {\em Physics Letters A}, {\bf 156},
192.

Chang, Z., and Yan, H.\ (1991a).\ {\em Physics Letters A}, {\bf 158}, 242.

Chang, Z., and Yan, H.\ (1991b).\ {\em Physical Review A}, {\bf 43}, 6043.

Chang, Z., and Yan, H.\ (1991c).\ {\em Physical Review A}, {\bf 44}, 7405.

Cornwell, J. F.\ (1984).\ {\em Group Theory in Physics}, Vol.\ 1, Academic
Press,
New York, p.\ 117.

Daskaloyannis, C.\ (1991).\ {\em Journal of Physics A}, {\bf 24}, L789.

Daskaloyannis, C.\ (1992).\ {\em Journal of Physics A}, {\bf 25}, 2261.

Eleonsky, V. M., and Korolev, V. G.\ (1995).\ {\em Journal of Physics A}, {\bf
28}, 4973.

Eleonsky, V. M., Korolev, V. G., and Kulagin, N. E.\ (1994).\ {\em Chaos}, {\bf
4}, 583.

Eleonsky, V. M., Korolev, V. G., and Kulagin, N. E.\ (1995).\ {\em Doklady
Rossiiskoi Akademii Nauk}, {\bf 342}, 1.

Fairlie, D. B., and Nuyts, J.\ (1994).\ {\em Journal of Mathematical Physics},
{\bf 35}, 3794.

Fairlie, D. B., and Zachos, C. K.\ (1991).\ {\em Physics Letters B}, {\bf 256},
43.

Fivel, D. I.\ (1990).\ {\em Physical Review Letters}, {\bf 65}, 3361.

Green, H. S.\ (1953).\ {\em Physical Review}, {\bf 90}, 270.

Greenberg, O. W.\ (1990).\ {\em Physical Review Letters}, {\bf 64}, 705.

Greenberg, O. W.\ (1991).\ {\em Physical Review D}, {\bf 43}, 4111.

Guichardet, A.\ (1998).\ {\em Journal of Mathematical Physics}, {\bf 39}, 4965.

Hayashi, T.\ (1990).\ {\em Communications in Mathematical Physics}, {\bf 127},
129.

Irac-Astaud, M., and Rideau, G.\ (1992).\ On the existence of quantum
bihamiltonian systems: The harmonic oscillator case, Universit\'e Paris VII
preprint, PAR-LPTM92.

Irac-Astaud, M., and Rideau, G.\ (1993).\ {\em Letters in Mathematical Physics},
{\bf 29}, 197.

Irac-Astaud, M., and Rideau, G.\ (1994).\ {\em Theorical and Mathematical
Physics}, {\bf 99}, 658.

Jannussis, A.\ (1993).\ {\em Journal of Physics A}, {\bf 26}, L233.

Jannussis, A., Brodimas, G., and Mignani, R.\ (1991).\ {\em Journal of
Physics A},
{\bf 24}, L775.

Jordan, T. F., Mukunda, N., and Pepper, S. V.\ (1963).\ {\em Journal of
Mathematical Physics}, {\bf 4}, 1089.

Katriel, J., and Quesne, C.\ (1996).\ {\em Journal of Mathematical Physics},
{\bf 37}, 1650.

Khare, A.\ (1992).\ {\em Journal of Physics A}, {\bf 25}, L749.

Khare, A.\ (1993).\ {\em Journal of Mathematical Physics}, {\bf 34}, 1277.

Khare, A., Mishra, A. K., and Rajasekaran, G.\ (1993a). {\em International
Journal of Modern Physics A}, {\bf 8}, 1245.

Khare, A., Mishra, A. K., and Rajasekaran, G.\ (1993b). {\em Modern Physics
Letters A}, {\bf 8}, 107.

Kosi\'nski, P., Majewski, M., and Ma\'slanka, P.\ (1997).\ {\em Journal of
Physics A}, {\bf 30}, 3983.

Kuryshkin, V.\ (1980).\ {\em Annales de la Fondation Louis de Broglie}, {\bf 5},
111.

McDermott, R. J., and Solomon, A. I.\ (1994).\  {\em Journal of Physics A}, {\bf
27}, L15.

Macfarlane, A. J.\ (1989).\ {\em Journal of Physics A}, {\bf 22}, 4581.

Macfarlane, A. J.\ (1994).\ {\em Journal of Mathematical Physics}, {\bf 35},
1054.

Man'ko, V. I., Marmo, G., Sudarshan, E. C. G., and Zaccaria, F.\ (1997).\
{\em Physica Scripta}, {\bf 55}, 528.

Meljanac, S., and Milekovi\'c, M.\ (1996).\ {\em International Journal of Modern
Physics A}, {\bf 11}, 1391.

Meljanac, S., Milekovi\'c, M., and Pallua, S.\ (1994).\ {\em Physics Letters B},
{\bf 328}, 55.

Mishra, A. K., and Rajasekaran, G.\ (1991a).\ {\em Pramana - Journal of
Physics},
{\bf 36}, 537.

Mishra, A. K., and Rajasekaran, G.\ (1991b).\ {\em Pramana - Journal of
Physics},
{\bf 37}, 455(E).

Ohnuki, Y., and Kamefuchi, S.\ (1982).\ {\em Quantum Field Theory and
Parastatistics}, Springer-Verlag, Berlin.

Paolucci, A., and Tsohantjis, I.\ (1997).\ {\em Physics Letters A}, {\bf 234},
27.

Plyushchay, M. S.\ (1996a).\ {\em Modern Physics Letters A}, {\bf 11}, 397.

Plyushchay, M. S.\ (1996b).\ {\em Annals of Physics (N.Y.)}, {\bf 245}, 339.

Polychronakos, A. P.\ (1992).\ {\em Physical Review Letters}, {\bf 69}, 703.

Quesne, C.\ (1994a).\ {\em Journal of Physics A}, {\bf 27}, 5919.

Quesne, C.\ (1994b).\ {\em Physics Letters A}, {\bf 193}, 245.

Quesne, C.\ (1995).\ {\em Modern Physics Letters A}, {\bf 10}, 1323.

Quesne, C., and Vansteenkiste, N.\ (1995).\ {\em Journal of Physics A}, {\bf
28}, 7019.

Quesne, C., and Vansteenkiste, N.\ (1996).\ {\em Helvetica Physica Acta}, {\bf
69}, 141.

Quesne, C., and Vansteenkiste, N.\ (1997).\ {\em Czechoslovak Journal of
Physics}, {\bf 47}, 115.

Quesne, C., and Vansteenkiste, N.\ (1998).\ {\em Physics Letters A}, {\bf 240},
21.

Quesne, C., and Vansteenkiste, N.\ (1999).\ {\em Helvetica Physica Acta}, {\bf
72}, 71.

Rideau, G.\ (1992).\ {\em Letters in Mathematical Physics}, {\bf 24}, 147.

Rubakov, V. A., and Spiridonov, V. P.\ (1988).\ {\em Modern Physics Letters A},
{\bf 3}, 1337.

Solomon, A. I.\ (1998).\ In {\em Fifth International Conference on Squeezed
States and Uncertainty Relations, Balatonfured, Hungary, May 27--31, 1997},
D.\ Han, J.\ Janszky, Y.\ S.\ Kim, and V.\ I.\ Man'ko, eds., NASA Conference
Publication 1998-206855, NASA Goddard Space Flight Center, Greenbelt, Maryland,
p.\ 157.

Sukhatme, U. P., Rasinariu, C., and Khare, A.\ (1997).\ {\em Physics Letters A},
{\bf 234}, 401.

Sun, C.-P., and Fu, H.-C.\ (1989).\ {\em Journal of Physics A}, {\bf 22}, L983.

Tsohantjis, I., Paolucci, A., and Jarvis, P. D.\ (1997).\ {\em Journal of
Physics A}, {\bf 30}, 4075.

Vasiliev, M. A.\ (1991).\ {\em International Journal of Modern Physics A}, {\bf
6}, 1115.

Veselov, A. P., and Shabat, A. B.\ (1993).\ {\em Funktsionalnyi Analiz i Ego
Prilozheniia}, {\bf 27}, 1.

Witten, E.\ (1981).\ {\em Nuclear Physics B}, {\bf 185}, 513.

Wolfes, J.\ (1974).\ {\em Journal of Mathematical Physics}, {\bf 15}, 1420.
%
%
\newpage

\begin{table}[h]

\caption{Classification of ${\cal A}^{(2)}(G(N))$ unirreps. Here $k_0$ may
take any  integer value.}

\vspace{1cm}
\begin{center}
\begin{tabular}{cccc}
  \hline\\[-0.2cm]
  Type & $n_0$ & $c$ & Conditions\\[0.2cm]
  \hline\\[-0.2cm]
  BFB & $2k_0$ & $n_0$ & $\alpha_0 > -1$\\[0.2cm]
  BFB & $2k_0 + 1$ & $n_0 + \alpha_0$ & $\alpha_0 < 1$\\[0.2cm]
  FD (d=1) & $2k_0$ & $n_0$ & $\alpha_0 = -1$\\[0.2cm]
  FD (d=1) & $2k_0 + 1$ & $n_0 + 1$ & $\alpha_0 = 1$\\[0.2cm]
  \hline
\end{tabular}
\end{center}

\end{table}
%
%
\newpage

\begin{table}[h]

\caption{Classification of ${\cal A}^{(3)}(G(N))$ unirreps. Here $k_0$ may take
any integer value.}

\vspace{1cm}
\begin{center}
\begin{tabular}{cccc}
  \hline\\[-0.2cm]
  Type & $n_0$ & $c$ & Conditions\\[0.2cm]
  \hline\\[-0.2cm]
  BFB & $3k_0$ & $n_0$ & $\alpha_0 > -1$, $\alpha_1 > -2 -\alpha_0$\\[0.2cm]
  BFB & $3k_0 + 1$ & $n_0 + \alpha_0$ & $\alpha_0 < 2$, $\alpha_1 > -1$\\[0.2cm]
  BFB & $3k_0 + 2$ & $n_0 + \alpha_0 + \alpha_1$ & $\alpha_0 < 1 - \alpha_1$,
        $\alpha_1 < 2$\\[0.2cm]
  FD (d=1) & $3k_0$ & $n_0$ & $\alpha_0 = -1$\\[0.2cm]
  FD (d=1) & $3k_0 + 1$ & $n_0 + \alpha_0$ & $\alpha_1 = -1$\\[0.2cm]
  FD (d=1) & $3k_0 + 2$ & $n_0 + 1$ & $\alpha_1 = 1 - \alpha_0$\\[0.2cm]
  FD (d=2) & $3k_0$ & $n_0$ & $\alpha_0 > -1$, $\alpha_1 = -2 -
\alpha_0$\\[0.2cm]
  FD (d=2) & $3k_0 + 1$ & $n_0 + 2$ & $\alpha_0 = 2$, $\alpha_1 > -1$\\[0.2cm]
  FD (d=2) & $3k_0 + 2$ & $n_0 + \alpha_0 + 2$ & $\alpha_0 < -1$, $\alpha_1
  = 2$\\[0.2cm]
  \hline
\end{tabular}
\end{center}

\end{table}
%
%
\newpage
\begin{table}[h]

\caption{Classification of ${\cal A}^{(4)}(G(N))$ unirreps. Here $k_0$ may
take any integer value.}

\vspace{1cm}
\begin{center}
\begin{tabular}{cccc}
  \hline\\[-0.2cm]
  Type & $n_0$ & $c$ & Conditions\\[0.2cm]
  \hline\\[-0.2cm]
  BFB & $4k_0$ & $n_0$ & $\alpha_0 > -1$, $\alpha_1 > -2 -\alpha_0$, $\alpha_2 >
         - 3 - \alpha_0 - \alpha_1$\\[0.2cm]
  BFB & $4k_0 + 1$ & $n_0 + \alpha_0$ & $\alpha_0 < 3$, $\alpha_1 > -1$,
         $\alpha_2 > - 2 - \alpha_1$\\[0.2cm]
  BFB & $4k_0 + 2$ & $n_0 + \alpha_0 + \alpha_1$ & $\alpha_0 < 2 - \alpha_1$,
        $\alpha_1 < 3$, $\alpha_2 > -1$\\[0.2cm]
  BFB & $4k_0 + 3$ & $n_0 + \alpha_0 + \alpha_1 + \alpha_2$ & $\alpha_0 < 1 -
        \alpha_1 - \alpha_2$, $\alpha_1 < 2 - \alpha_2$, $\alpha_2 < 3$\\[0.2cm]
  FD (d=1) & $4k_0$ & $n_0$ & $\alpha_0 = -1$\\[0.2cm]
  FD (d=1) & $4k_0 + 1$ & $n_0 + \alpha_0$ & $\alpha_1 = -1$\\[0.2cm]
  FD (d=1) & $4k_0 + 2$ & $n_0 + \alpha_0 + \alpha_1$ & $\alpha_2 = -1$\\[0.2cm]
  FD (d=1) & $4k_0 + 3$ & $n_0 + 1$ & $\alpha_2 = 1 - \alpha_0 -
        \alpha_1$\\[0.2cm]
  FD (d=2) & $4k_0$ & $n_0$ & $\alpha_0 > -1$, $\alpha_1 = -2 -
\alpha_0$\\[0.2cm]
  FD (d=2) & $4k_0 + 1$ & $n_0 + \alpha_0$ & $\alpha_1 > -1$, $\alpha_2 = - 2 -
        \alpha_1$\\[0.2cm]
  FD (d=2) & $4k_0 + 2$ & $n_0 + 2$ & $\alpha_1 = 2 - \alpha_0$, $\alpha_2
        > -1$\\[0.2cm]
  FD (d=2) & $4k_0 + 3$ & $n_0 + \alpha_0 + 2$ & $\alpha_0 < -1$, $\alpha_2
        = 2 - \alpha_1$\\[0.2cm]
  FD (d=3) & $4k_0$ & $n_0$ & $\alpha_0 > -1$, $\alpha_1 > -2 - \alpha_0$,
        $\alpha_2 = - 3 - \alpha_0 - \alpha_1$\\[0.2cm]
  FD (d=3) & $4k_0 + 1$ & $n_0 + 3$ & $\alpha_0 = 3$, $\alpha_1 > -1$,
        $\alpha_2 > - 2 - \alpha_1$\\[0.2cm]
  FD (d=3) & $4k_0 + 2$ & $n_0 + \alpha_0 + 3$ & $\alpha_0 < -1$, $\alpha_1
        = 3$, $\alpha_2 > -1$\\[0.2cm]
  FD (d=3) & $4k_0 + 3$ & $n_0 + \alpha_0 + \alpha_1 + 3$ & $\alpha_0 < - 2 -
        \alpha_1$, $\alpha_1 < -1$, $\alpha_2 = 3$\\[0.2cm]
  \hline		
\end{tabular}
\end{center}

\end{table}

\end{document}